\documentclass[onecolumn,nobibnotes,nofootinbib,superscriptaddress]{revtex4}
\usepackage{amsmath,amssymb,bm}
\usepackage[a4paper,bindingoffset=0.2in,left=0.8in,right=0.8in,top=1in,bottom=1in,footskip=.25in]{geometry}
\usepackage{graphicx,subfigure,epsfig,epstopdf}
\usepackage{bigints}
\usepackage{color}
\usepackage[colorlinks=true,linkcolor=blue,anchorcolor=blue,citecolor=blue,urlcolor=blue]{hyperref}

\newcommand{\be}{\begin{equation}}
\newcommand{\ee}{\end{equation}}
\newcommand{\bea}{\begin{eqnarray}}
\newcommand{\eea}{\end{eqnarray}}
\newcommand{\benn}{\begin{displaymath}}
\newcommand{\eenn}{\end{displaymath}}
\newcommand{\beann}{\begin{eqnarray*}}
\newcommand{\eeann}{\end{eqnarray*}}

\begin{document}

\title{Inclusive diffractive heavy quarkonium photoproduction including quark subprocesses}
\author{Jiayu Wu}
\email{wujyedu@gmail.com}
\affiliation{Department of Physics, Guizhou University, Guiyang, 550025, China}
\author{Yanbing Cai}
\email{myparticle@163.com}
\affiliation{Guizhou Key Laboratory in Physics and Related Areas, Guizhou University of Finance and Economics, Guiyang 550025, China}
\author{Wenchang Xiang}
\email{wxiangphy@gmail.com}
\affiliation{Department of Physics, Guizhou University, Guiyang, 550025, China}
\affiliation{Guizhou Key Laboratory in Physics and Related Areas, Guizhou University of Finance and Economics, Guiyang 550025, China}


\begin{abstract}
The inclusive $J/\Psi$, $\Psi(2S)$ and $\Upsilon(1S)$ direct and resolved photoproduction are investigated by including the quark subprocesses in the framework of non-relativistic quantum chromodynamics (NRQCD). We find that the theoretical total cross section of heavy quarkonium productions are in good agreement with the data available at HERA, once the $\gamma q$, $qg$ and $qq$ subprocesses in the heavy quark pair productions are taken into account. The inclusive diffractive rapidity and transverse momentum distributions of $J/\Psi$, $\Psi(2S)$ and $\Upsilon(1S)$ in $pp$, $pPb$ and $PbPb$ collisions at LHC are also studied by our quark improved NRQCD model combined with the resolved pomeron model. We find that the contributions from the quark involved subprocesses can reach to $8\%$ in the rapidity distribution and $6\%$ in the transverse momentum distribution. The numerical results show that the contributions from quark involved subprocesses are significant in heavy quarkonium photoprodution.
\end{abstract}

\maketitle


\section{Introduction}
\label{sec:intro}
Since the discovery of the $J/\Psi$ charmonium in 1974\cite{Aubert,Augustin}, the quarkonium production has attracted much attention experimentally and theoretically as it provides an excellent way to investigate both the perturbation and non-perturbation regimes of quantum chromodynamics. In recent years, the measurements released by the ALICE Collaboration\cite{Aamodt}, CMS Collaboration\cite{Silvestre}, ATLAS Collaboration\cite{Aad} and LHCb Collaboration\cite{Aaij} provide an unique opportunity to explore the heavy quarkonium production mechanism. On the theoretical aspect, several approaches have been proposed to calculate the production of heavy quarkonium, such as the color singlet mechanism (CSM)\cite{Petrelli,Braaten,Einhorn,Chang,Carlson}, the color dipole mechanism\cite{Song,Kharzeev,Goncalves,Qiu}, the color-evaporation model (CEM)\cite{Godbole,Fujii}, $k_T$-factorization method\cite{Gribov,Kniehl1,Baranov1} and the NRQCD\cite{Bodwin,Mathews,He1,He2,Baranov2,Kang1,Kang2,Butenschoen,Han}. Among these approaches, the most widely used methods for heavy quarkonia production are CSM and NRQCD.

To describe the heavy quarkonium production, a key ingredient is the nonperturbative evolution of an intermediate $Q\bar{Q}$ pair into the final quarkonium.  The CSM is the first model to describe this nonperturbative evolution. The CSM was successful in describing quarkonium production at low energy\cite{Schuler}. However, there are considerable discrepancies between CSM predictions and the experimental data at Tevatron. The CSM prediction of inclusive $J/\Psi$ hadroproduction underestimates the experimental data by more than one order of magnitude  \cite{Abe}. For inclusive $\Psi(2S)$ hadroproduction, the discrepancy is even larger, reaching a factor of 50\cite{Abe}.  The reason for these discrepancies is the coupled quantum number of the heavy quark pair is assumed to be in a color singlet in the CSM. To solve this puzzle, Bodwin, Braaten and Lepage proposed the NRQCD mechanism by introducing the color octet process\cite{Bodwin}. In the NRQCD framework, the quarkonium can evolve from color singlet intermediate heavy quark pair, as well as color octet heavy quark pair, which greatly fill the gap between the CSM predictions and experimental measurements\cite{Kniehl2}. The theoretical results about quarkonium hadroproduction from NRQCD are also in good agreement with the experimental measurements of $\chi_{cJ}$ and $\Upsilon(1S)$\cite{Kramer}. As NRQCD factorization method has achieved great success in description heavy quarkonuim hadroproduction, many efforts have been done to investigate the heavy quarkonuim photoproduction with NRQCD.

Photoproduction process is a vital process in electron-proton deep inelastic scattering (DIS) and high energy ultra-peripheral hadron collisions, since it offers significant information about the interaction mechanisms\cite{Cai1}. The $J/\Psi$ photoproduction at HERA has been studied in the framework of the NRQCD by taking into account the color octet contributions. It was found that the color octet matrix elements, which are obtained by fitting the hadroproduction data at the Tevatron, overestimate the HERA data an order of magnitude\cite{Ko}. This excess is disappeared when the higher order effects are included\cite{Kniehl2}. The quarkonium photoproduction cross section at LHC has been investigated by using the CSM in Ref.\cite{Goncalves2}. Meanwhile, the NRQCD has been used to study the inclusive diffractive quarkonium photoproduction at LHC by combining with the resolved pomeron model\cite{Goncalves3}. Later on, the $\eta_c$ production has been performed in inclusive and diffractive processes with the NRQCD model in Ref.\cite{Goncalves4}. Although considerable efforts have been made to improve the performance of the theoretical models, the theoretical calculations can match with the experimental data under certain uncertainties, the mechanism for heavy quarkonium production is still far from the deep understanding. Therefore, a comprehensive analysis of the contributions from different subprocesses may be helpful for understanding the heavy quarkonium production mechanism, such as $g+g\rightarrow Q\bar{Q}+g$\cite{Yang}.

In our previous study, the inclusive diffractive heavy quarkonium photoproduction has been investigated by taking into account the $g+g\rightarrow Q\bar{Q}+g$\cite{Yang}. We found that the resolved photoproduction processes have a significant contribution to the heavy quarkonium production. In this paper, the quark subprocesses, $\gamma + q \rightarrow Q\bar{Q}+q$, $g + q \rightarrow Q\bar{Q}+q$, $q + \bar{q} \rightarrow Q\bar{Q}+g$, are included into the direct and resolved photoproduction processes on top of our previous studies in Ref.\cite{Yang}. We find that the contributions from the quark involved subprocesses can reach to $8\%$ in the rapidity distribution and $6\%$ in the transverse momentum distribution.

\section{Inclusive diffractive quarkonium photoproduction in NRQCD }

In this section, we firstly introduce the mechanism of the diffractive heavy quarkonium photoproduciton based on the NRQCD combined with the resolved pomeron model. Then we give the total cross section for heavy quarkonium produciton in direct and resolved photoproduction processes. The total cross section and the rapidity distribution of the heavy quarkonium photoproduction in $pp$, $pPb$ and $PbPb$ collisions are presented at the end of this section.

\subsection{The mechanism of the diffractive quarkonium photoproduciton}
\label{sec:nrqcd}
The NRQCD is one of the most widely used theoretical tool to describe the quarkonium production\cite{Bodwin,Mathews,He1,He2,Baranov2,Kang1,Kang2,Butenschoen,Han}. In the NRQCD formalim, the cross section for the production of a heavy quarkonium $H$ can be factorized as\cite{Hao}

\be
\sigma(ab\rightarrow H+X)=\sum_n\sigma(ab\rightarrow Q\overline{Q}[n]+X)<\mathcal{O}_{[n]}^H> ,
\label{nrqcd}
\ee
where $\sigma(ab\rightarrow Q\overline{Q}[n]+X)$ is the short distance process and $<O_{[n]}^H>$ is the long distance process. In the short distance process, a pair of quark-antiquark is produced at collision point, and the quark-antiquark forms a special system $Q\overline{Q}[n]$. In the long distance process, the special system form a bound state, and eventually the bound state evolutes into the final quarkonium through soft gluon emission. In the NRQCD formulism, the short distance matrix elements can be calculated perturbatively, but the long distance matrix is a non perturbative process and it is usually given by the lattice QCD calculation or extracting from experimental data.

\begin{figure}[h]
\setlength{\unitlength}{1.5cm}
\begin{center}
\epsfig{file=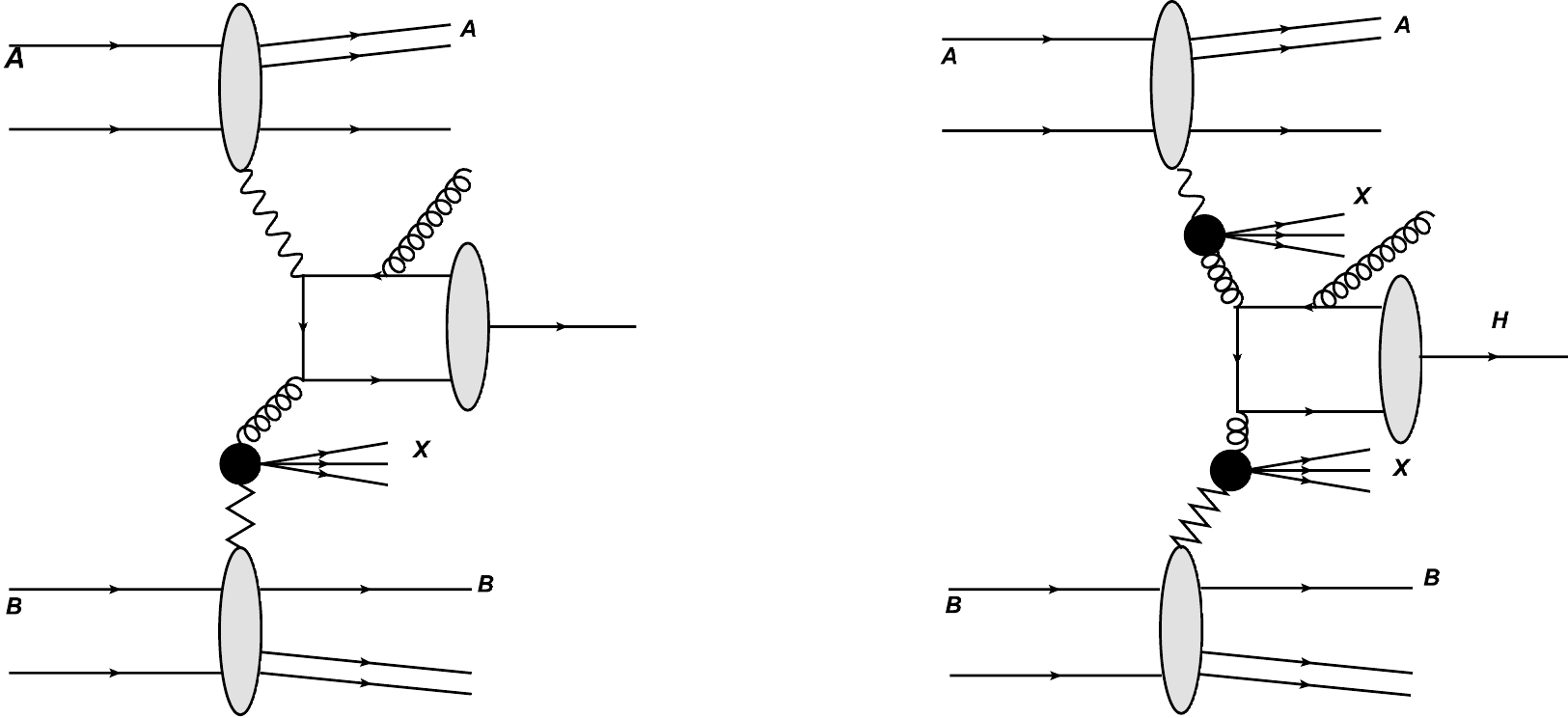, width=12cm,height=5cm}
\end{center}
\caption{The schematic diagrams of inclusive diffractive heavy quarkonium direct (left panel) and resolved (right panel) photoproduction in the resolved pomeron model.}
\label{fig1}
\end{figure}

In this work, we combine the NRQCD factorization approach with resolved pomeron model to investigate the inclusive diffractive heavy quarkonium photoproduction in hadron-hadron collisions. The photoproduction has direct and resolved two type processes according to the scattering mechanism of the photon. A schematic diagrams are shown in Fig.\ref{fig1}. In the direct photoproduction processes (Fig.\ref{fig1} (left panel)), the photons emitted by the hadron ($A$) interact with the partons (quarks and gluons) from the resolved pomerons in the hadron ($B$) to produce final particle $H$. In the resolved photoproduction processes (Fig.\ref{fig1} (right panel)), due to the Heisenberg uncertainty principle the energetic photons from the energetic hadron ($A$) can fluctuate into partons. Then these partons interact with the partons from the resolved pomerons to form the final heavy quarkonium.


\subsection{The cross section of the direct photoproduciton}
\label{sec:dir}
According to the NRQCD formulism, the total cross section can be factorized into the partonic distribution function and differential cross section. Therefore, the total cross section of direct photoproduction process $\gamma + B\rightarrow H + X$ can be written as

\be
\sigma_{dir}(\gamma+B\rightarrow H+X)=\int dzdp^2_T\frac{x_{dir}f_{b/B}(x_{dir},Q^2)}{z(1-z)}\frac{d\sigma}{dt}(\gamma+b\rightarrow H+X),
\label{totdir}
\ee
where $f_{b/B}(x_{dir},Q^2)$ is the diffractive parton distribution\cite{Ingelman}. The $x_{dir}$ in Eq.(\ref{totdir}) is the momentum fraction of hadron carried by the quark or gluon and can be written as

\be
x_{dir}=\frac{p^2_T+M^2(1-z)}{W^2_{\gamma h}z(1-z)},
\label{xdir}
\ee
where $p_T$ is the transverse momentum of the quarkonium and $M$ is the mass of the quarkonium. The $W_{\gamma h}$ in Eq.(\ref{xdir}) is the photon-hadron center-of-mass energy
\be
W_{\gamma h}=\sqrt{2\omega\sqrt{s}},
\label{cms}
\ee
where $\sqrt{s}$ represents the center-of-mass energy of the $AB$ collision system and $z$ is the fraction of the photon energy. The $\omega$ in Eq.(\ref{cms}) is the photon energy in the proton-proton collision system
\be
\omega=\frac{M}{2}\exp(\pm y).
\label{ey}
\ee
Note that the fraction of photon energy carried by the quarkonium in Eq.(\ref{totdir}) is integrated over the range\cite{Aaron},
\be
0.3\le z\le 0.9.
\ee

\subsection{The cross section of resolved photoproduction}
\label{res}
In the resolved photoproduction process, the energetic photons fluctuate into quarks and gluons which then interact with the partons from the pomeron. According to the factorization formalism, the cross section for the resolved photoproduction process is given by

\be
\sigma_{res}(a+B \rightarrow H+X)=\int dzdp^2_Tdx\frac{x_{res}f_{a/\gamma}(x_{res},Q^2)f_{b/B}(x,Q^2)}{z(1-\frac{z}{x_{res}})}\frac{d\sigma}{dt}(a+b\rightarrow H+X),
\label{totres}
\ee
where $f_{a/\gamma}(x_{res},Q^2)$ is the parton distribution function of the hadron-like photon\cite{Gluck}, and $x_{res}$ is the momentum fraction of photon carried by the quark or gluon

\be
x_{res}=\frac{xp^2_T+M^2(x-z)}{W^2_{\gamma p}z(1-\frac{z}{x})}.
\label{res}
\ee

In the early studies, the heavy quarkonium was investigated in the CSM\cite{Goncalves2,Berger,Baier1,Krmer}, in which the $Q\overline{Q}$ produced by the partonic process is color-singlet. In direct and resolved photoproduction, the color-singlet processes are given by

\be
\gamma+g\rightarrow Q\overline{Q}[^3S_1,1]+g,
\label{res9}
\ee
\be
g+g\rightarrow Q\overline{Q}[^3S_1,1]+g.
\label{res10}
\ee
The contribution of $\gamma g$ process to heavy quarkonium photoproduction was studied in Ref.\cite{Goncalves2,Goncalves3} and the contribution of $gg$ process was investigated in our previous studies in Ref.\cite{Yang}. Although the theoretical calculations can match with the experimental data under certain uncertainties, there are many sources that contribute to the theoretical uncertainties, such as the uncertainties in long-distance matrix elements, scale, and parton distribution functions\cite{Yang}. To get a comprehensive understanding about the mechanism of heavy quarkonium production, an analysis of contribution from the quark involved subprocesses is essential. In fact that the quarks are also active partons in hadron-like photons and pomerons, so one has to include the contributions from quark processes in order to improve our understanding of the heavy quarkonium production mechanism. In this work, we take into account the quark processes to investigate the inclusive diffractive heavy quarkonium photoproduction by combining the NRQCD model with resolved pomeron model. We call this approach as quark improved NRQCD model.  When the quark contributions are considered, the following processes should be included
\be
\gamma+q\rightarrow Q\overline{Q}[^3S_1,1]+q,
\label{res11}
\ee
\be
g+q\rightarrow Q\overline{Q}[^3S_1,1]+q,
\label{res12}
\ee
\be
q+\overline{q}\rightarrow Q\overline{Q}[^3S_1,1]+g,
\label{res13}
\ee
where the above three subprocesses can be obtained in Ref.\cite{Klasen}.
We know that the CSM can be used to describe $J/\Psi$ production in $\gamma p$ or $ep$ collisions. However, the theoretical prediction based on the CSM underestimates the cross section of inclusive $J/\Psi$ production measured in $pp$ collisions\cite{Abe}, and this inconsistency cannot be attributed to different energy scales, the heavy quark mass or parton distribution function\cite{Cano-Coloma}. To solve this difficulty, Bodwin, Braaten, and Lepage\cite{Bodwin} developed the Color-Octet mechanism (COM) based on NRQCD which allows a systematic calculation of inclusive heavy quarkonium production cross sections. In the COM, the processes can be expressed as follows\cite{Klasen}

\be
\gamma+g\rightarrow Q\overline{Q}[^3S_0, ^3S_1, ^3P_{0,1,2},8]+g,
\label{res14}
\ee
\be
g+g\rightarrow Q\overline{Q}[^1S_0, ^3S_1, ^3P_{0,1,2},8]+g,
\label{res15}
\ee
\be
\gamma+q\rightarrow Q\overline{Q}[^3S_0, ^3S_1, ^3P_{0,1,2},8]+q,
\label{res16}
\ee
\be
g+q\rightarrow Q\overline{Q}[^1S_0, ^3S_1, ^3P_{0,1,2},8]+q,
\label{res17}
\ee
\be
q+\overline{q}\rightarrow Q\overline{Q}[^1S_0, ^3S_1, ^3P_{0,1,2},8]+g.
\label{res18}
\ee
Using Eqs.(\ref{res9})-(\ref{res18}), we can get the contributions of heavy quarkonium in the direct and resolved photoproduction processes.

\subsection{The total cross section and the rapidity distribution of the heavy quarkonium photoproduction}
\label{sec:tot}
In this subsection, we will firstly discuss the total cross section of heavy quarkonium production. Then we give the rapidity distribution of the quarkonium photoproduction on the basis of total cross section. The total cross section for the heavy quarkonium in the photoproduction process is given by\cite{Cai2}

\be
\sigma_{tot}(A+ B\rightarrow A\otimes H+X\otimes B) =\int d\omega\frac{dN_{\gamma/A}(\omega)}{d\omega}\sigma_{\gamma B}\rightarrow HX\otimes B\\
+\int d\omega\frac{dN_{\gamma/B}(\omega)}{d\omega}\sigma_{\gamma A}\rightarrow HX\otimes A,
\label{totcs}
\ee
where $\otimes$ denotes the existence of a rapidity gap in the final state and $\omega$ denotes the energy of photon. The $A$ and $B$ denote a hadron or nucleus. For $pp$ collision, using the relationship between rapidity and energy in Eq.(\ref{ey}), the rapidity distribution can be written as

\be
\frac{d\sigma_{tot}^{pp\rightarrow pH+Xp}}{dy}=[\omega\frac{dN^p_\gamma}{d\omega}\sigma^{\gamma p\rightarrow HXp}]_{\omega=\omega_l}+[\omega\frac{dN^p_\gamma}{d\omega}\sigma^{\gamma p\rightarrow HXp}]_{\omega=\omega_r},
\label{totpp}
\ee
where the subscrips $l(r)$ denote photon flux from the right(left) proton.
The equivalent photon flux $\frac{dN_\gamma^p}{d\omega}$ of the relativistic proton in Eq.(\ref{totpp}) is given by\cite{Klein,Drees}

\be
\frac{dN_\gamma^p}{d\omega}=\frac{\alpha_{em}}{2\pi\omega}[1+(1-\frac{2\omega}{\sqrt{s}})^2](\ln\Omega-\frac{11}{6}+\frac{3}{\Omega}-\frac{3}{2\Omega^2}+\frac{1}{3\Omega^3}),
\label{tot}
\ee
with

\be
\Omega=1+[(0.71\rm {GeV}^2)/Q_{min}^2],
\label{tot}
\ee
where $Q_{min}^2 = \omega^2/[\gamma_L^2(1-2\omega/\sqrt{s})]\approx(\omega/\gamma_L)^2$ and $\gamma_L$ is the lorentz factor.

To calculate the rapidity distribution in Eq.(\ref{totpp}), we need to specify the diffractive parton distribution $f_{b/p}(x, Q^2)$, which can be expressed as the convolution of the pomeron flux in the proton $f_\mathcal{P}^P(x_\mathcal{P})$ and parton distribution in the pomeron\cite{Ingelman}

\be
f_{b/p}(x,Q^2)=\int_x^1\frac{dx_\mathcal{P}}{x_\mathcal{P}}f^p_\mathcal{P}(x_\mathcal{P})g_\mathcal{P}(\frac{x}{x_\mathcal{P}},Q^2),
\label{pdfp}
\ee
where the pomeron fluxes is given by
\be
f_\mathcal{P}^p(x_\mathcal{P})=\int_{t_{min}}^{t_{max}}dtf_\mathcal{P}^p(x_\mathcal{P},t)=\int_{t_{min}}^{t_{max}}\frac{\lambda e^{\beta t}}{x_\mathcal{P}^{2\alpha_{\mathcal{P}}(t)-1}}dt,
\label{tot}
\ee
with

\be
t_{max}=-m_p^2x_\mathcal{P}^2/(1-x_\mathcal{P}),
\label{res}
\ee
\be
t_{min}=-1\,\rm{GeV}^2.
\label{tot}
\ee
In Eq.(\ref{pdfp}), $g_\mathcal{P}(\frac{x}{x_\mathcal{P}}, Q^2)$ is the parton distribution in pomeron, which is given by a parameterized formula in Ref.\cite{Aktas}.

For $pPb$ collisions, the rapidity distribution is
\be
\frac{d\sigma_{tot}^{pPb\rightarrow pHXPb}}{dy}=[\omega\frac{dN^{Pb}_\gamma}{d\omega}\sigma^{\gamma Pb\rightarrow HXp}]_{\omega=\omega_l}+[\omega\frac{dN_\gamma^p}{d\omega}\sigma^{\gamma p\rightarrow HXPb}]_{\omega=\omega_r},
\label{totppb}
\ee
where $\frac{dN_\gamma^{Pb}}{d\omega}$ denotes the photon flux of lead\cite{Papageorgiu}

\be
\frac{dN_\gamma^{Pb}}{d\omega}=\frac{2Z^2\alpha_{em}}{\pi\omega}[\xi K_0(\xi)K_1(\xi)-\frac{\xi^2}{2}(K_1^2(\xi)-K_0^2(\xi))],
\label{tot}
\ee
with $\xi = \omega(R_p + R_{Pb})/\gamma_L$. The $K_{0}$ and $K_{1}$ are the modified Bessel functions and $Z$ is atomic number. Since the photon flux of the nuclei is larger than the photon flux of the proton, the rapidity distribution should be obvious asymmetry with a great peak at the lead side in the $pPb$ collision. The diffractive parton distribution of the lead can be expressed as follows\cite{Basso}
\be
f_{b/Pb}(x_\mathcal{P})=\int_{t_{min}}^{t_{max}}dtf_\mathcal{P}^{Pb}(x_\mathcal{P},t)=R_gA^2\int_{t_{min}}^{t_{max}}\frac{\lambda e^{\beta t}}{x_\mathcal{P}^{2\alpha_{\mathcal{P}}(t)-1}}dtF_A^2(t),
\label{tot}
\ee
where $R_{g}$ is the suppression factor associated to the nuclear shadowing and $F_A(t)$ is the nuclear form factor.

In addition to the rapidity distribution of heavy quarkonium production in $pp$ and $pPb$ collisions, we also study the heavy quarkonium photoproduction in the $PbPb$ collision by including the quark involved subprocesses as mentioned above. The rapidity distribution of $PbPb$ collision can be written as

\be
\frac{d\sigma_{tot}^{PbPb\rightarrow PbHXPb}}{dy}=[\omega\frac{dN^{Pb}_\gamma}{d\omega}\sigma^{\gamma Pb\rightarrow HXPb}]_{\omega=\omega_l}+[\omega\frac{dN_\gamma^{Pb}}{d\omega}\sigma^{\gamma Pb\rightarrow HXPb}]_{\omega=\omega_r}.
\label{totpbpb}
\ee

\section{Results}
\label{sec:numsol}
In this section, we firstly calculate the total cross section distribution of the inclusive $J/\Psi$, $\Psi(2S)$ and $\Upsilon(1S)$ production to explore the contribution of the quark subprocesses, and compare our theoretical calculations with the $J/\Psi$ experimental data from H1 Collaboration\cite{Aaron}. Then we use the quark improved NRQCD model to make predictions of rapidity and transverse momentum distributions of the inclusive diffractive $J/\Psi$, $\Psi(2S)$ and $\Upsilon(1S)$ photoproduction in $pp$, $pPb$ and $PbPb$ collisions.

In our studies, the masses of charm quark and bottom quark are taken 1.5 GeV and 4.5 GeV, respectively. The minimum value of the transverse momentum is set to 1 GeV. We set the factorization scale as $Q^2 = P^2_T + (M/2)^2$. In addition, the differential cross section of the partonic processes is calculated in Ref.\cite{Klasen}, and the long-distance matrix elements are listed in Table.1\cite{Yu}.

\begin{table}[h]
\begin{tabular}{cccc}
\toprule
$(\mathrm{GeV}^3)$ &$J/\Psi$ &$\Psi(2S)$ &$\Upsilon(1S)$\\
\hline
$<\mathcal{O}[^3S_1^{[1]}]>$& 1.2& 0.76&10.9\\
$<\mathcal{O}[^3S_0^{[8]}]>$& $0.018\pm0.0087$&$0.0080\pm 0.0067$&$0.121\pm0.00400$ \\
$<\mathcal{O}[^3S_1^{[8]}]>$& $0.0013\pm0.0013$& $0.00330\pm0.0021$& $0.0477\pm0.0334$\\
$<\mathcal{O}[^3P_0^{[8]}]>$& $(0.018\pm0.0087)m_c^2$&$(0.0080\pm 0.0067)m_c^2$&$(0.121\pm0.00400)m_b^2$ \\
$<\mathcal{O}[^3P_1^{[8]}]>$& $3\times(0.018\pm0.0087)m_c^2$&$3\times(0.0080\pm 0.0067)m_c^2$&- \\
$<\mathcal{O}[^3P_2^{[8]}]>$& $5\times(0.018\pm0.0087)m_c^2$&$5\times(0.0080\pm 0.0067)m_c^2$&- \\
\hline
\end{tabular}
\caption{Values of long-distance matrix elements for $J/\Psi$, $\Psi(2S)$ and $\Upsilon(1S)$.}
\end{table}

\begin{figure}[b]
\setlength{\unitlength}{1.5cm}
\begin{center}
\epsfig{file=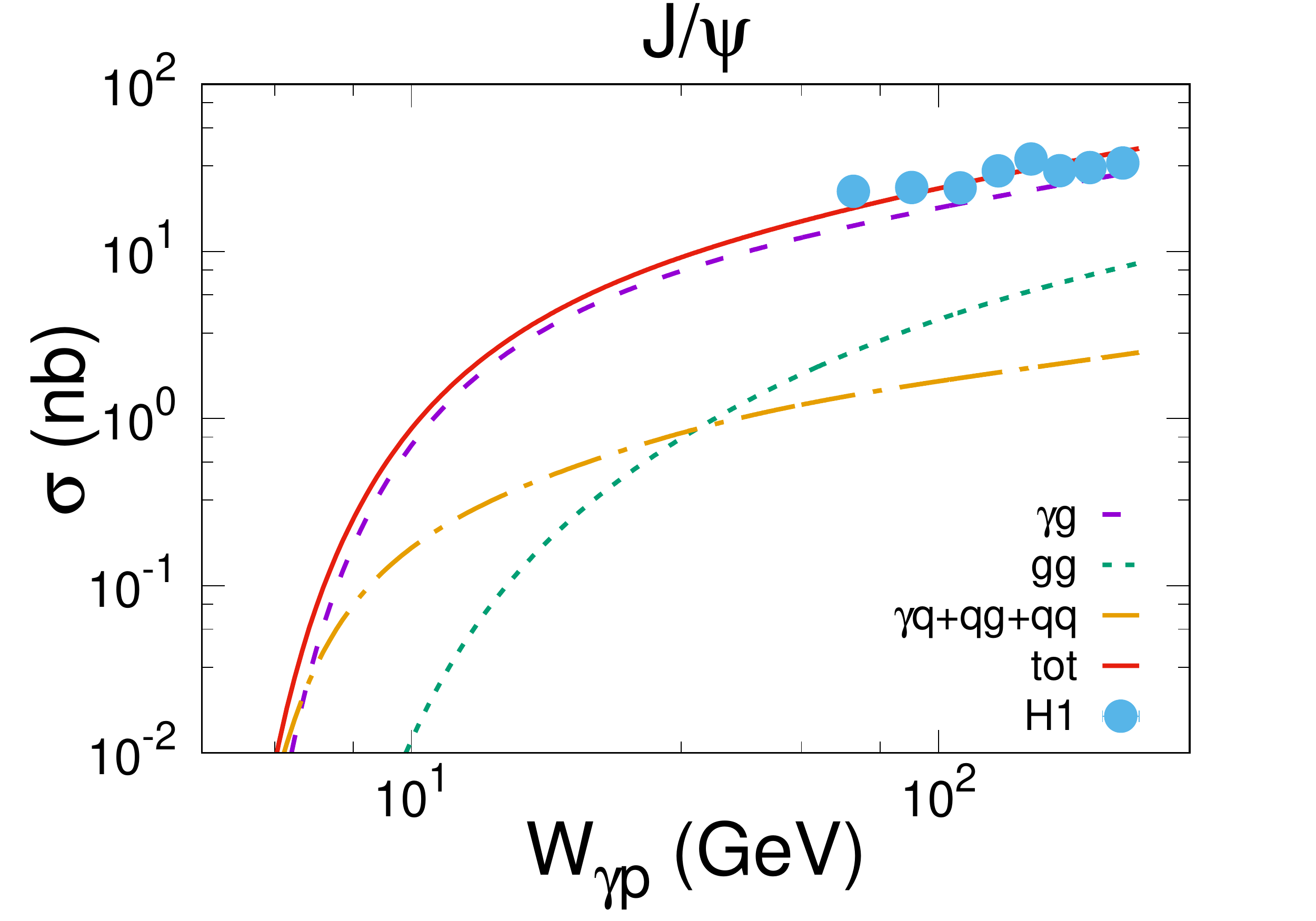, width=8cm, height=10cm,angle=270}
\end{center}
\caption{The energy dependence of the cross section for the $J/\Psi$ photoproduction in inclusive $\gamma p$ interactions. Data are from H1 Collaboration\cite{Aaron}.}
\label{fig2}
\end{figure}

In Fig.\ref{fig2}, we give the theoretical results for energy dependence of the inclusive $J/\Psi$ total cross section calculated by Eqs.(\ref{totdir}) and (\ref{totres}) with a parton direct from the proton and compare with the experimental data. The dash purple line is the contribution from the $\gamma g$ direct photoproduction process. The dotted green line is the result from the $gg$ resolved photoproduction process. The dot-dash orange line is the contributions from the processes of quarks($\gamma q$, $qq$, $qg$), which are introduced, for the first time, into theoretical calculations of the heavy quarkonium photoproduction in the NRQCD framework. The solid red line is the total cross section. As one can see that the quark improved NRQCD model can give a rather successful description of the $J/\Psi$ photoproduction data from H1 Collaboration, which indicates that the quark involved three subprocesses have a significant contributions.

\begin{figure}[t]
\setlength{\unitlength}{1.5cm}
\begin{center}
\epsfig{file=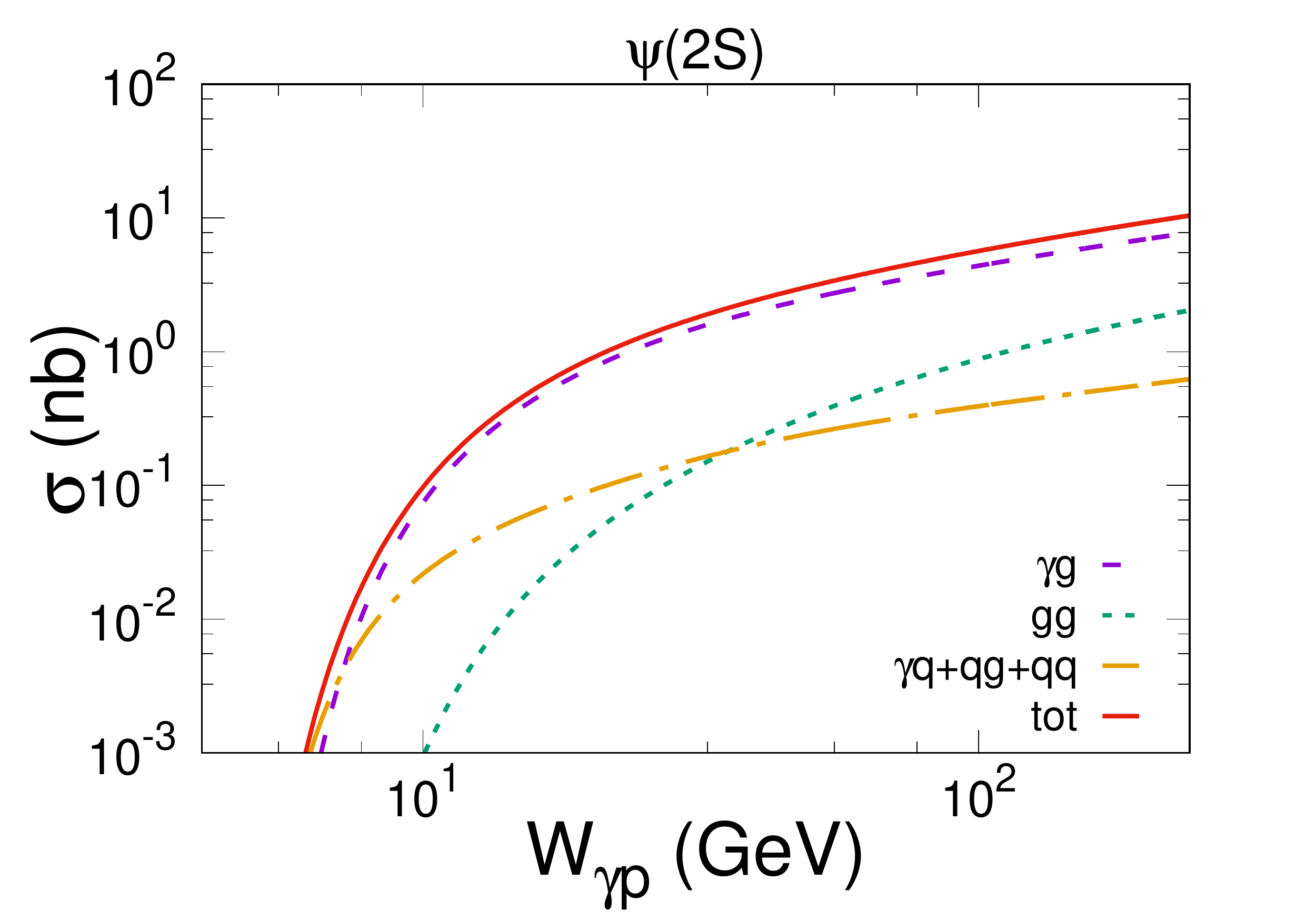, width=6cm,height=8cm,angle=270}
\epsfig{file=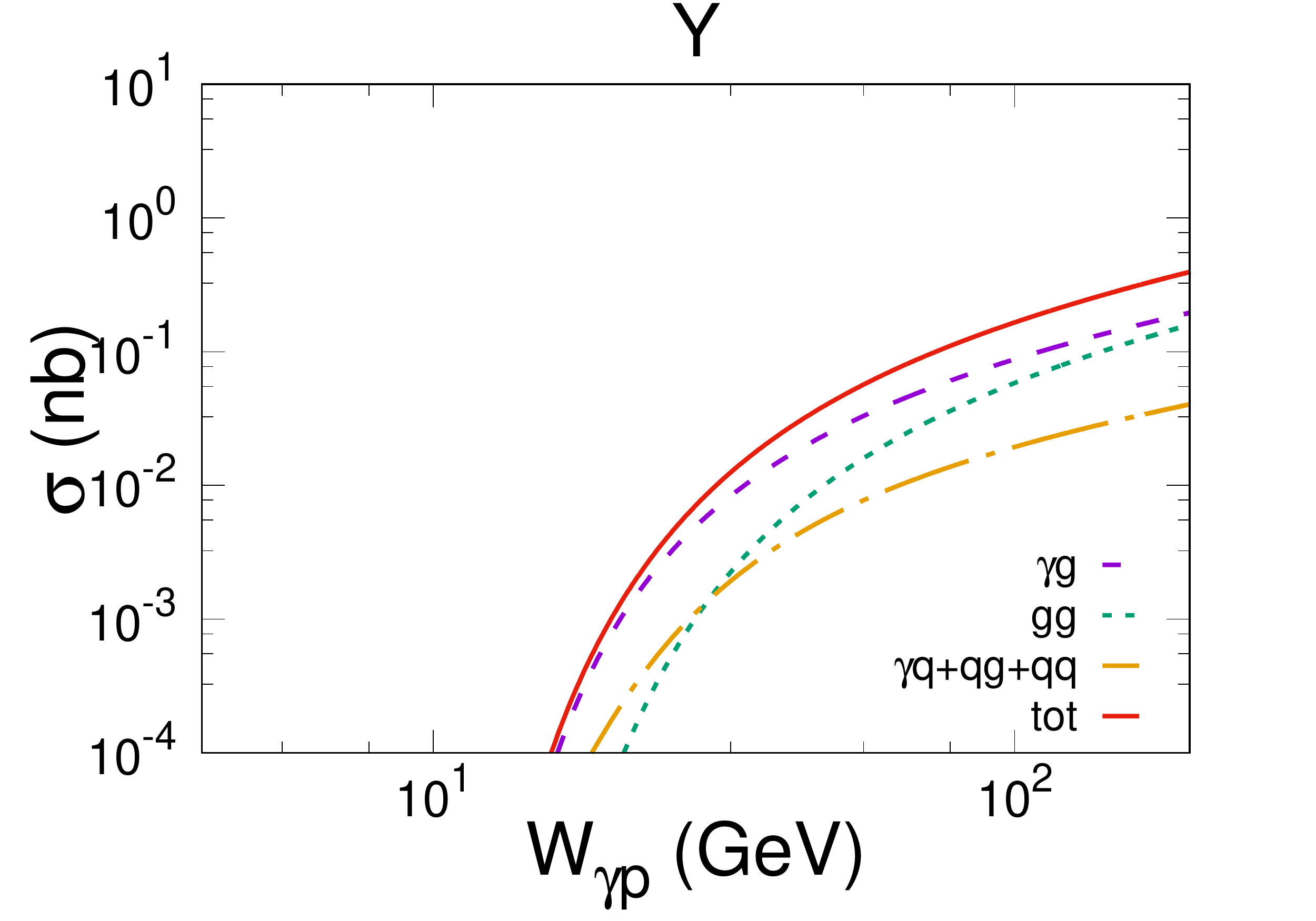, width=6cm,height=8cm,angle=270}
\end{center}
\caption{Predictions for the energy dependence of the cross section for the $\Psi(2S)$ and $\Upsilon(1S)$ photoproduction in inclusive $\gamma p$ interaction.}
\label{fig3}
\end{figure}

Using Eqs.(\ref{totdir}) and (\ref{totres}), the predictions of the inclusive total cross section for $\Psi(2S)$ and $\Upsilon(1S)$ are computed and shown in Fig.\ref{fig3}. We can see that the quark involved three subprocesses also have a significant contributions to the $\Psi(2S)$ and $\Upsilon(1S)$ productions. From Figs.\ref{fig2} and \ref{fig3}, we can see that in the low region of center-of-mass (or large $x$), the contributions of the quark involved three subprocesses are larger than the ones from $gg$ process, since the quark distribution is larger than the gluon distribution function in the low energy region.

\begin{figure}[t]
\setlength{\unitlength}{1.5cm}
\begin{center}
\epsfig{file=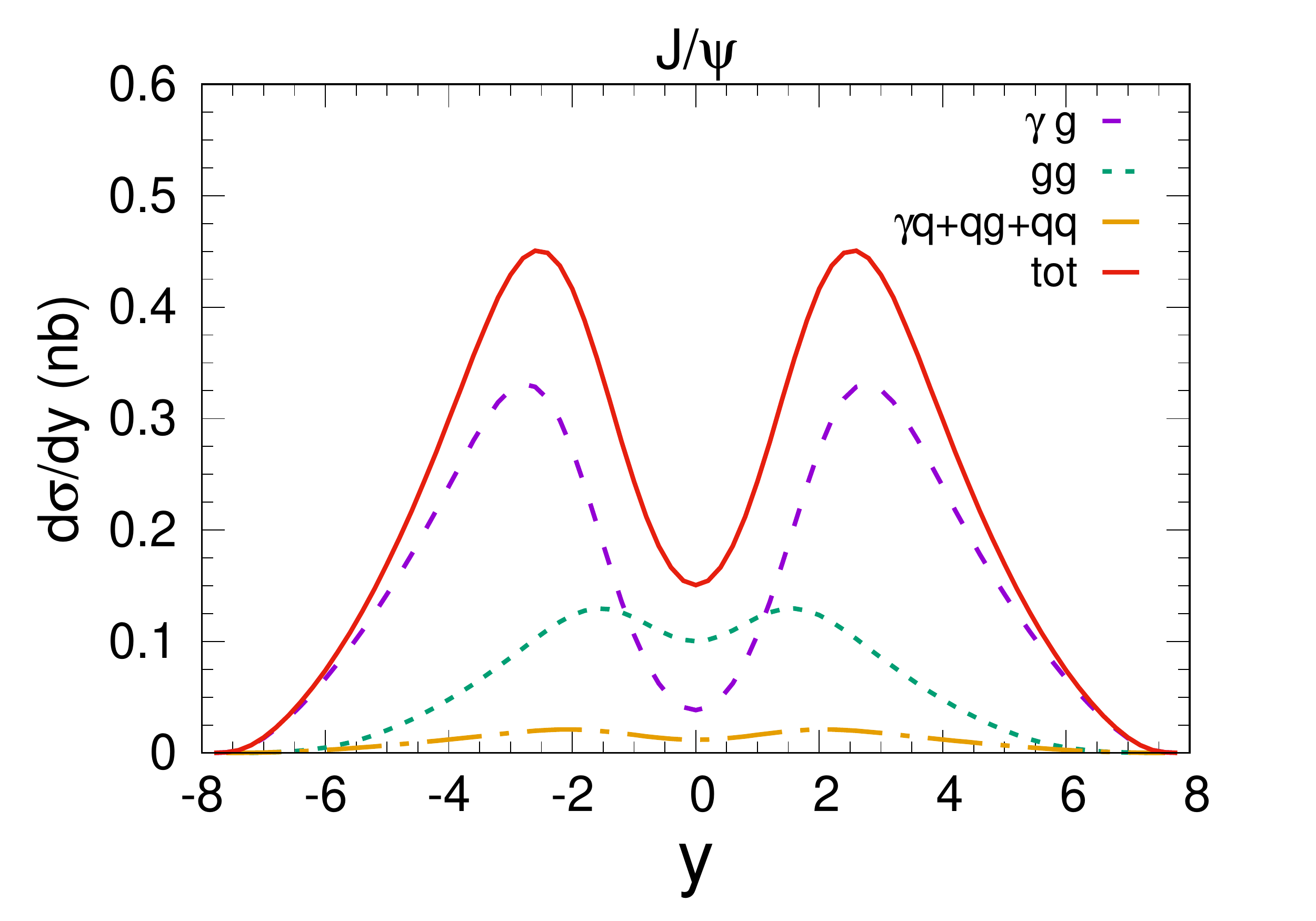, width=4cm,height=5cm,angle=270}
\epsfig{file=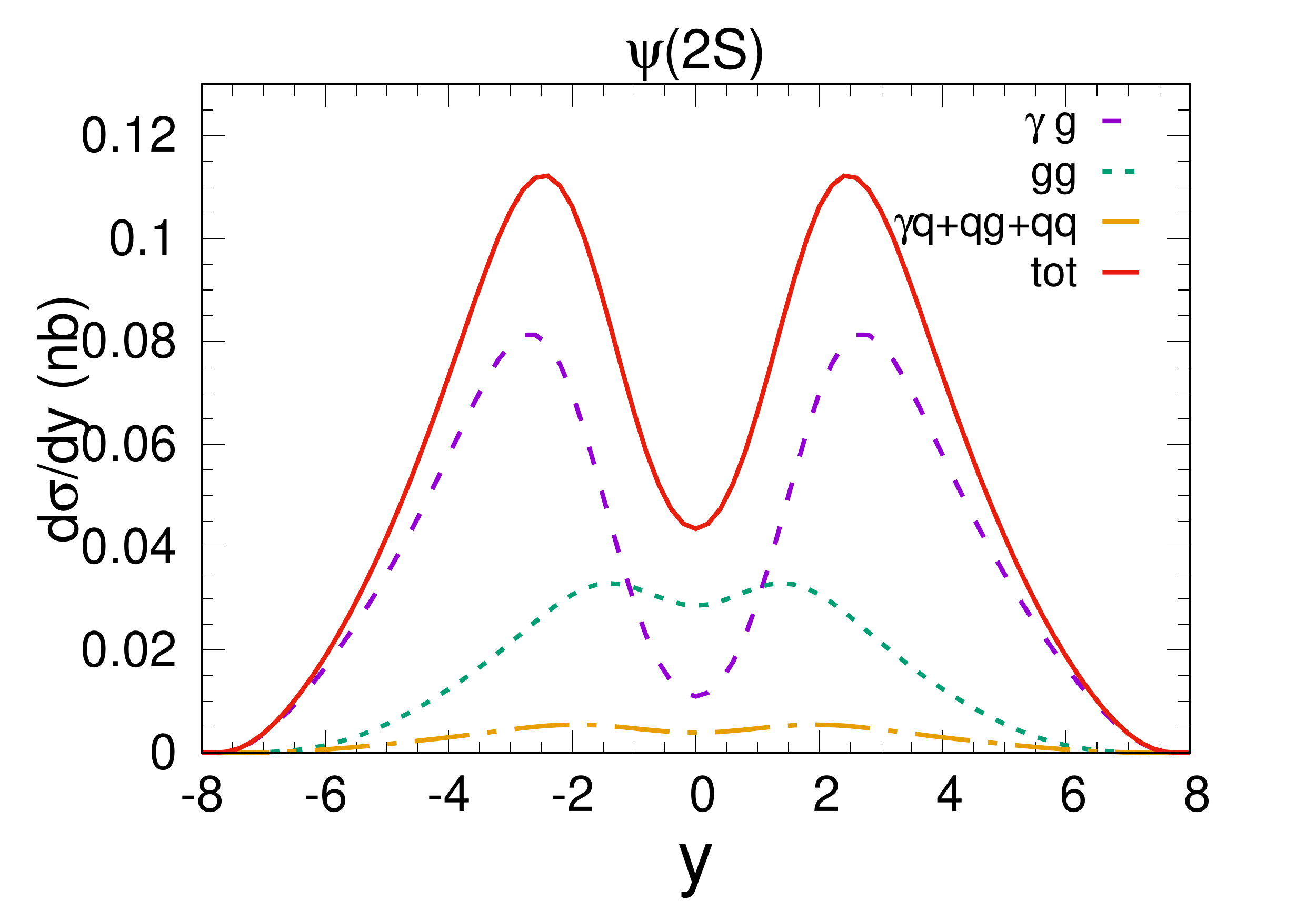, width=4cm,height=5cm,angle=270}
\epsfig{file=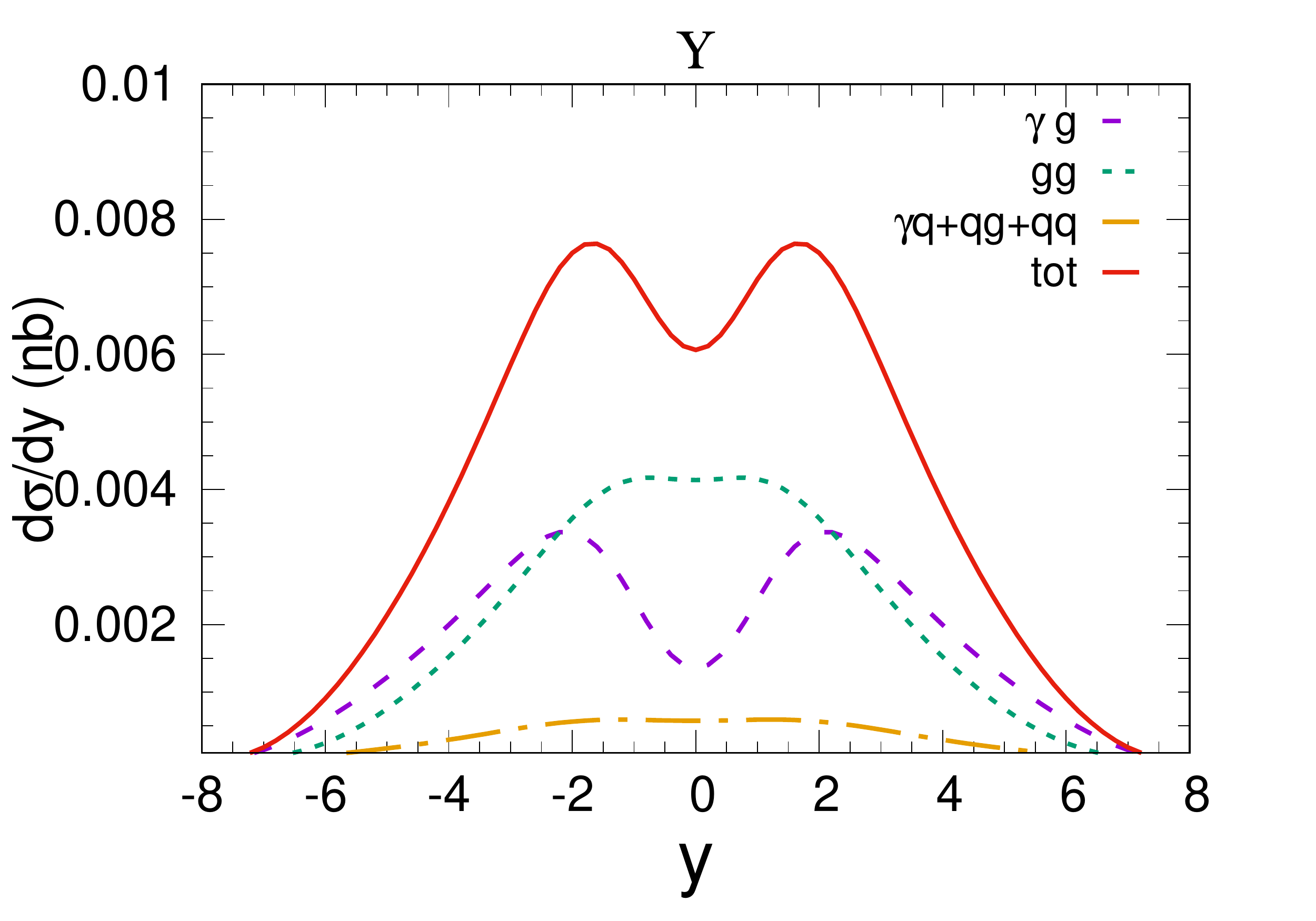, width=4cm,height=5cm,angle=270}
\end{center}
\caption{Rapidity distributions for the inclusive diffractive $J/\Psi$, $\Psi(2S)$ and $\Upsilon(1S)$ photoproduction in $pp$ collisions at $\sqrt{s}$ = 5.02 TeV.}
\label{fig4}
\end{figure}

From the inclusive $J/\Psi$, $\Psi(2S)$ and $\Upsilon(1S)$ photoproduction at HERA energies, one can see that the quark involved three subprocesses have a significant contributions to the total cross section. It is reasonable to believe that the three subprocesses mentioned above are also important in the heavy quarkonium production at LHC energies. So, we use the quark improved NRQCD model to study the heavy quarkonium photoproduction at LHC energies. Via using Eq.(\ref{totpp}), we calculate the predictions of the rapidity distribution for the inclusive diffractive $J/\Psi$, $\Psi(2S)$ and $\Upsilon(1S)$  in $pp$ collision at $\sqrt{s} = 5.02$ TeV, the relevant numerical results are shown in Fig.\ref{fig4}. We estimate the percentage of contributions from the quark involved three subprocesses, and find that the three subprocesses can reach to $8\%$ of the total rapidity distribution. Thus, one needs to include their contributions into the heavy quarkonium production for an accurate description of the experimental data.

\begin{figure}[t]
\setlength{\unitlength}{1.5cm}
\begin{center}
\epsfig{file=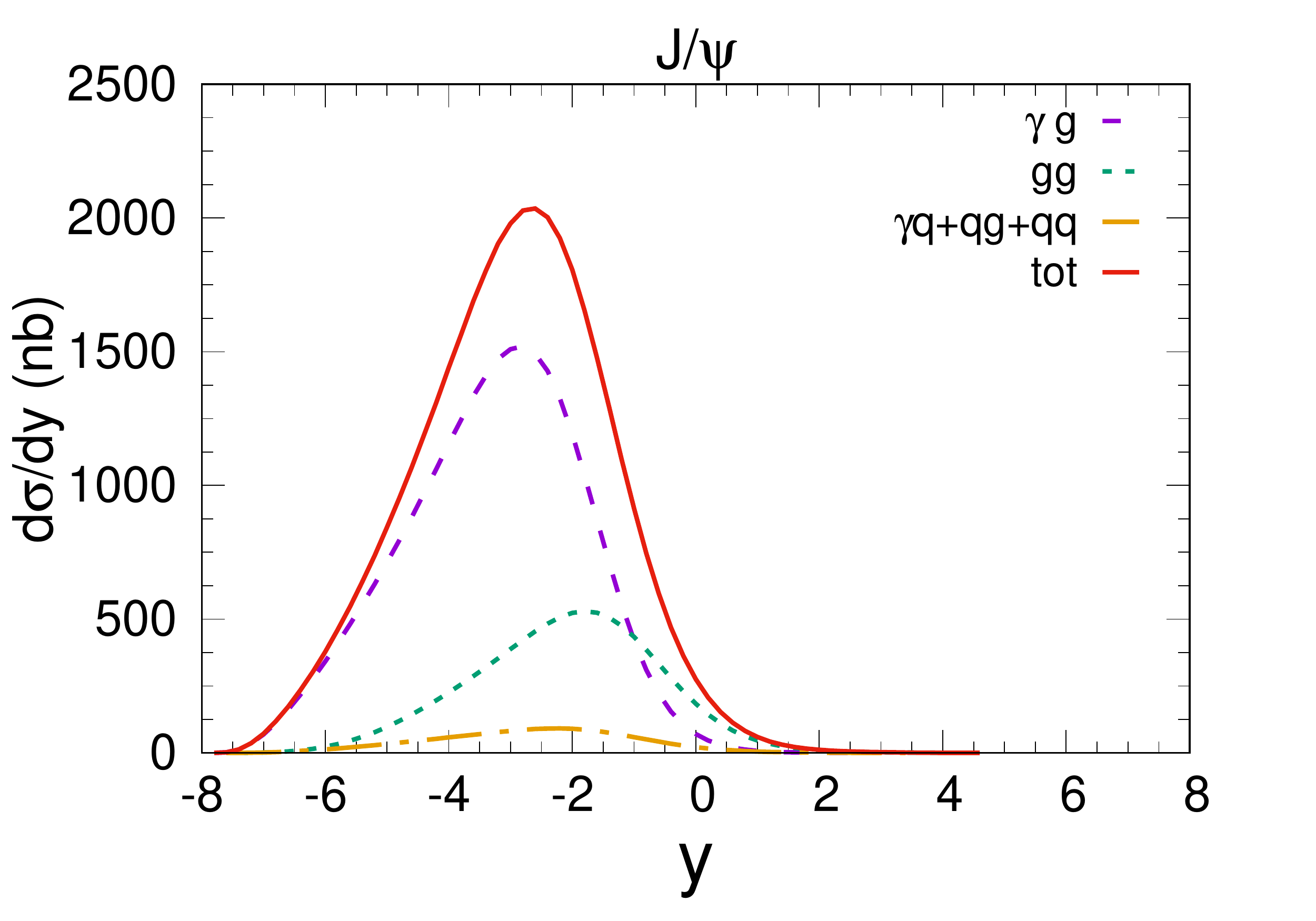, width=4cm,height=5cm,angle=270}
\epsfig{file=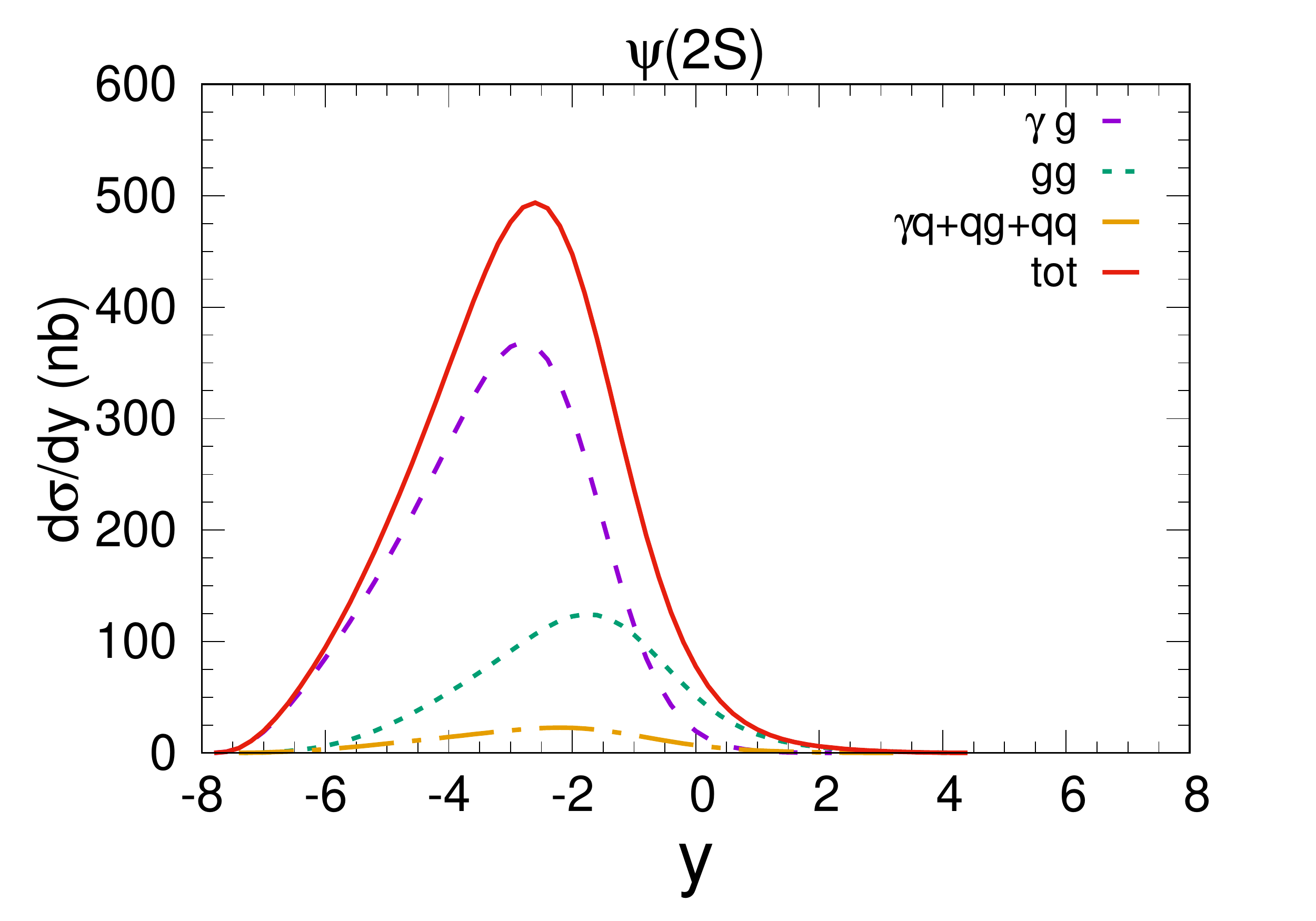, width=4cm,height=5cm,angle=270}
\epsfig{file=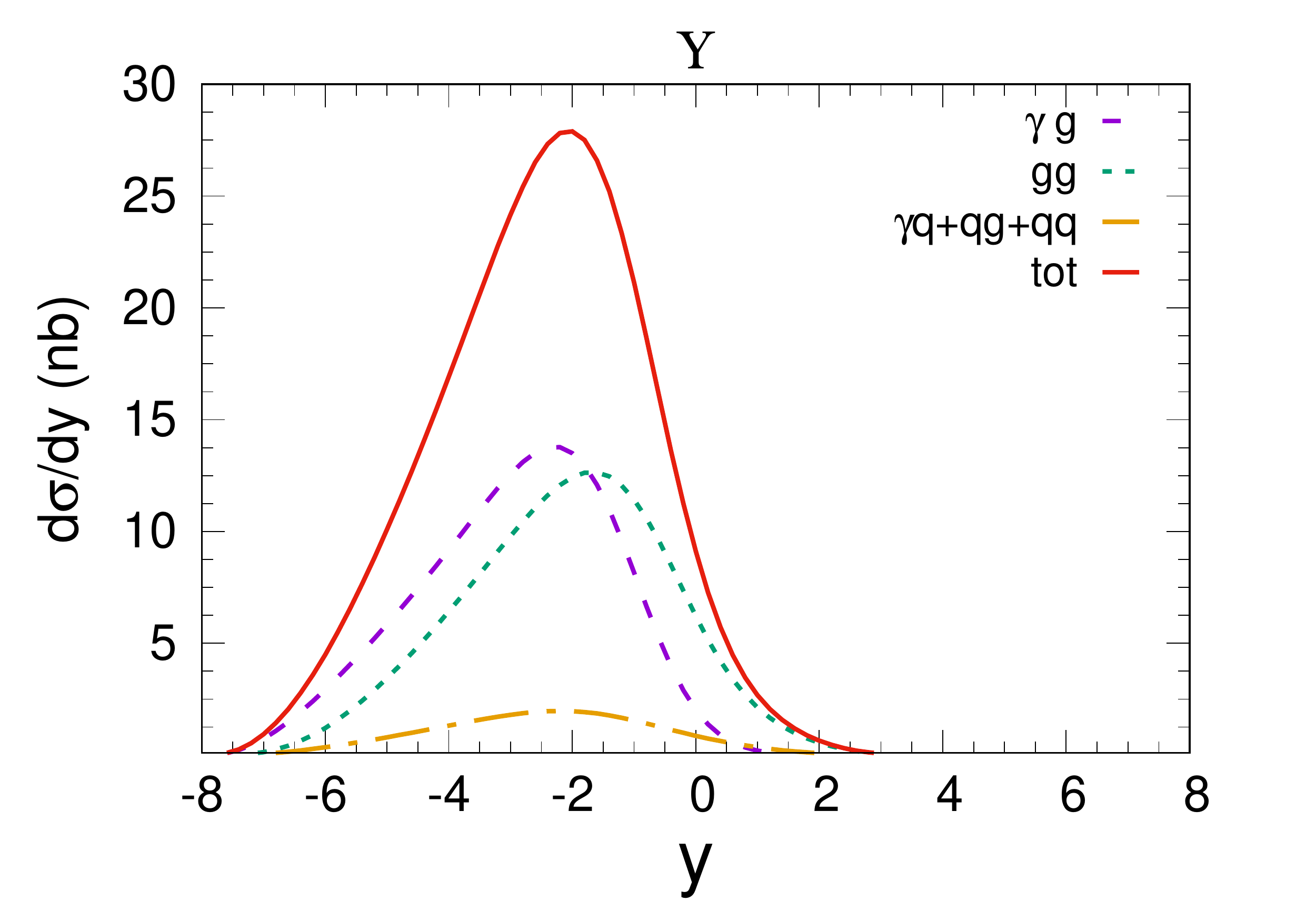, width=4cm,height=5cm,angle=270}
\end{center}
\caption{Rapidity distributions for the inclusive diffractive $J/\Psi$, $\Psi(2S)$ and $\Upsilon(1S)$ photoproduction in $pPb$ collision at $\sqrt{s}$ = 5.02 TeV.}
\label{fig5}
\end{figure}

\begin{figure}[t]
\setlength{\unitlength}{1.5cm}
\begin{center}
\epsfig{file=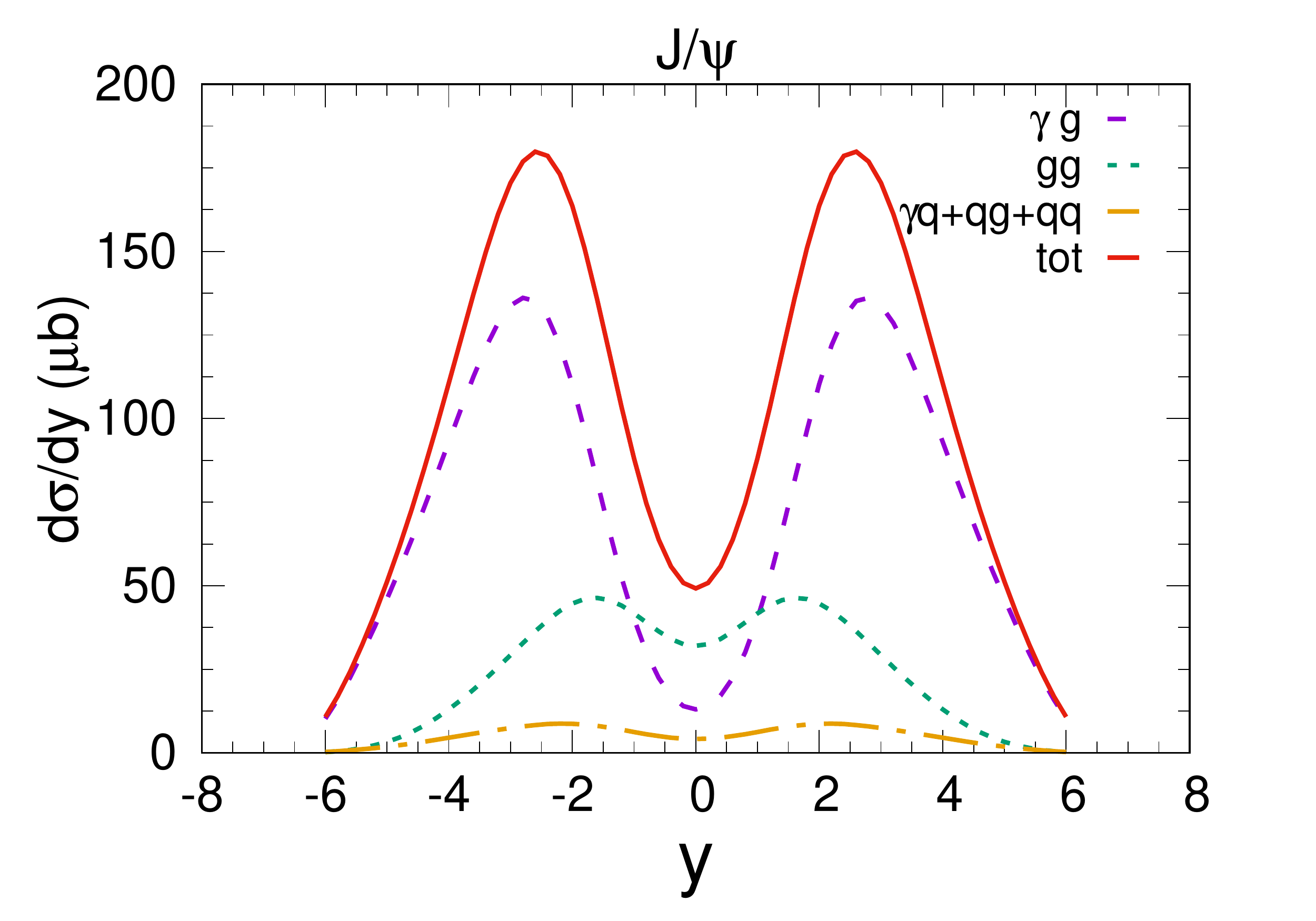, width=4cm,height=5cm,angle=270}
\epsfig{file=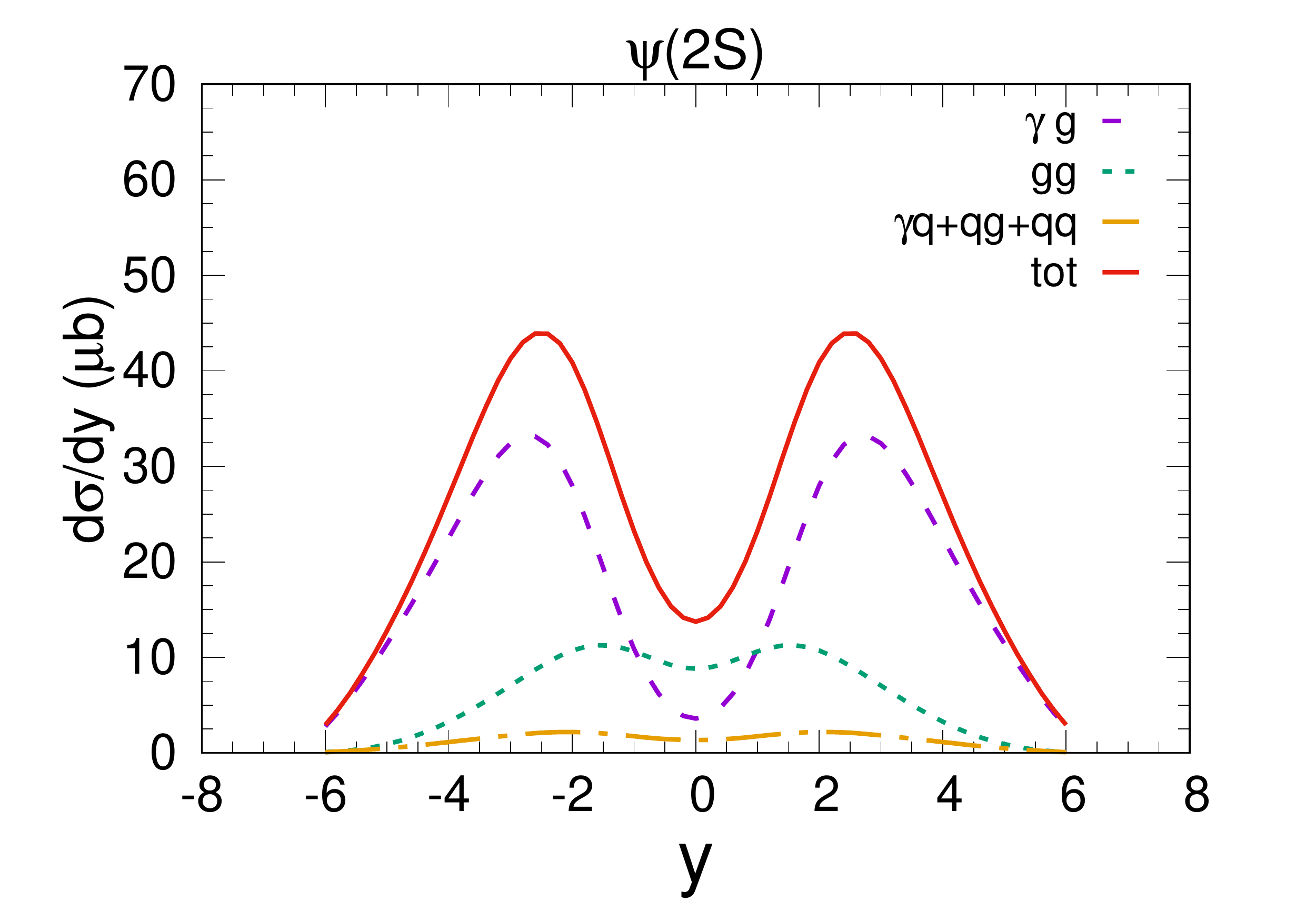, width=4cm,height=5cm,angle=270}
\epsfig{file=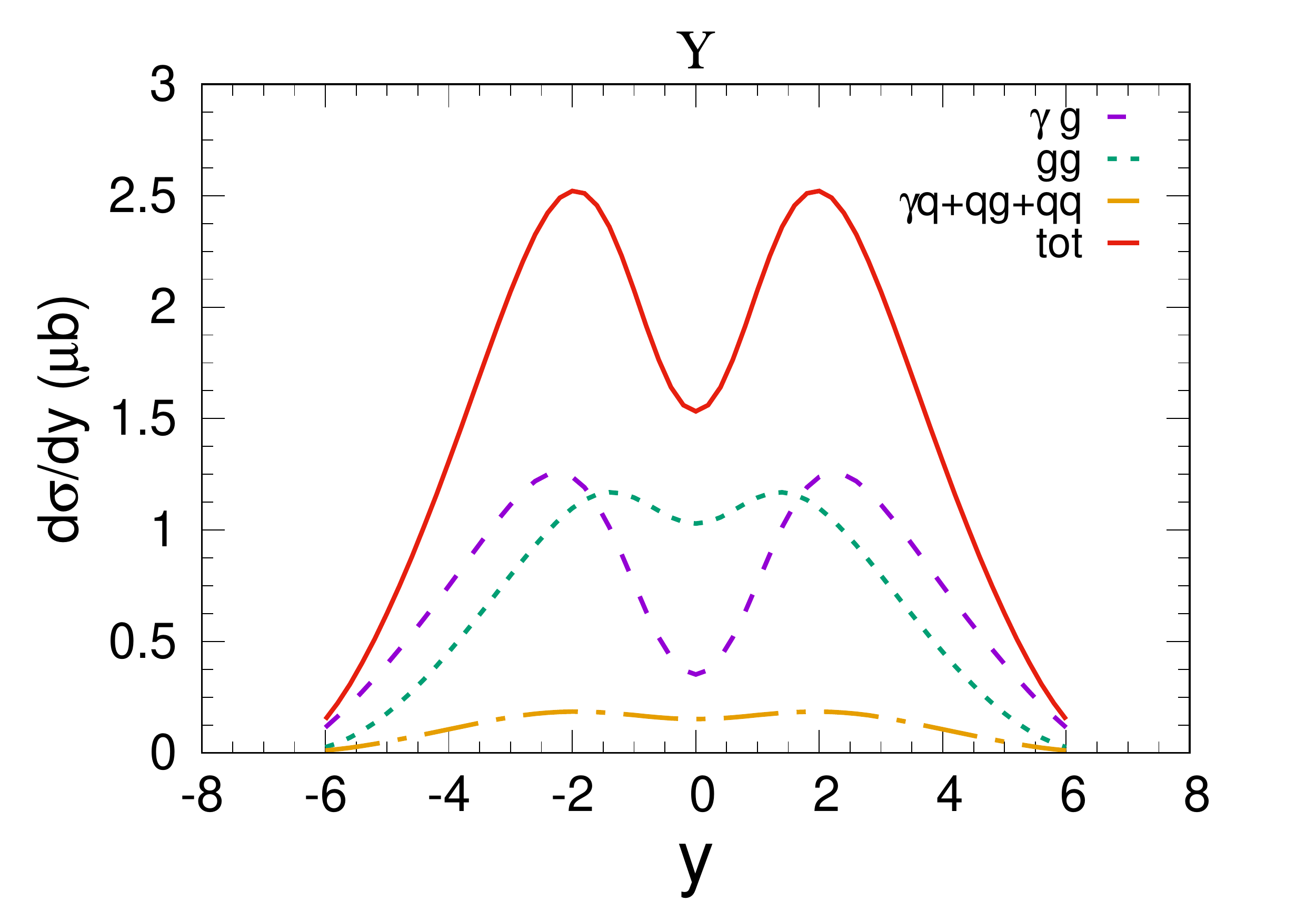, width=4cm,height=5cm,angle=270}
\end{center}
\caption{Rapidity distributions for the inclusive diffractive $J/\Psi$, $\Psi(2S)$ and $\Upsilon(1S)$ photoproduction in $PbPb$ collision at $\sqrt{s}$ = 5.02 TeV.}
\label{fig6}
\end{figure}

In the LHC energies, $pPb$ and $PbPb$ collisions have also attracted much attention. So we calculate the predictions of rapidity distributions for these two collision processes via Eqs.(\ref{totppb}) and (\ref{totpbpb}), respectively. We present the numerical calculations of the rapidity distributions for the inclusive diffractive $J/\Psi$, $\Psi(2S)$ and $\Upsilon(1S)$ photoproduciton in $pPb$ collision at $\sqrt{s} = 5.02$ TeV in Fig.\ref{fig5}. As expected, in $pPb$ collision the rapidity distributions are asymmetric, with the peak at the lead side, since the photon spectrum of a nucleus is much larger than a proton.

The rapidity distributions for the inclusive diffractive $J/\Psi$, $\Psi(2S)$ and $\Upsilon(1S)$ photoproduciton in $PbPb$ collision at $\sqrt{s} = 5.02$ TeV are shown in Fig.\ref{fig6}. As in the $pp$ collision, the total rapidity distributions are symmetric about the midrapidity. However, the rapidity distributions are larger than the $pp$ collision due to the enhancement of the photon flux.

\begin{figure}[b]
\setlength{\unitlength}{1.5cm}
\begin{center}
\epsfig{file=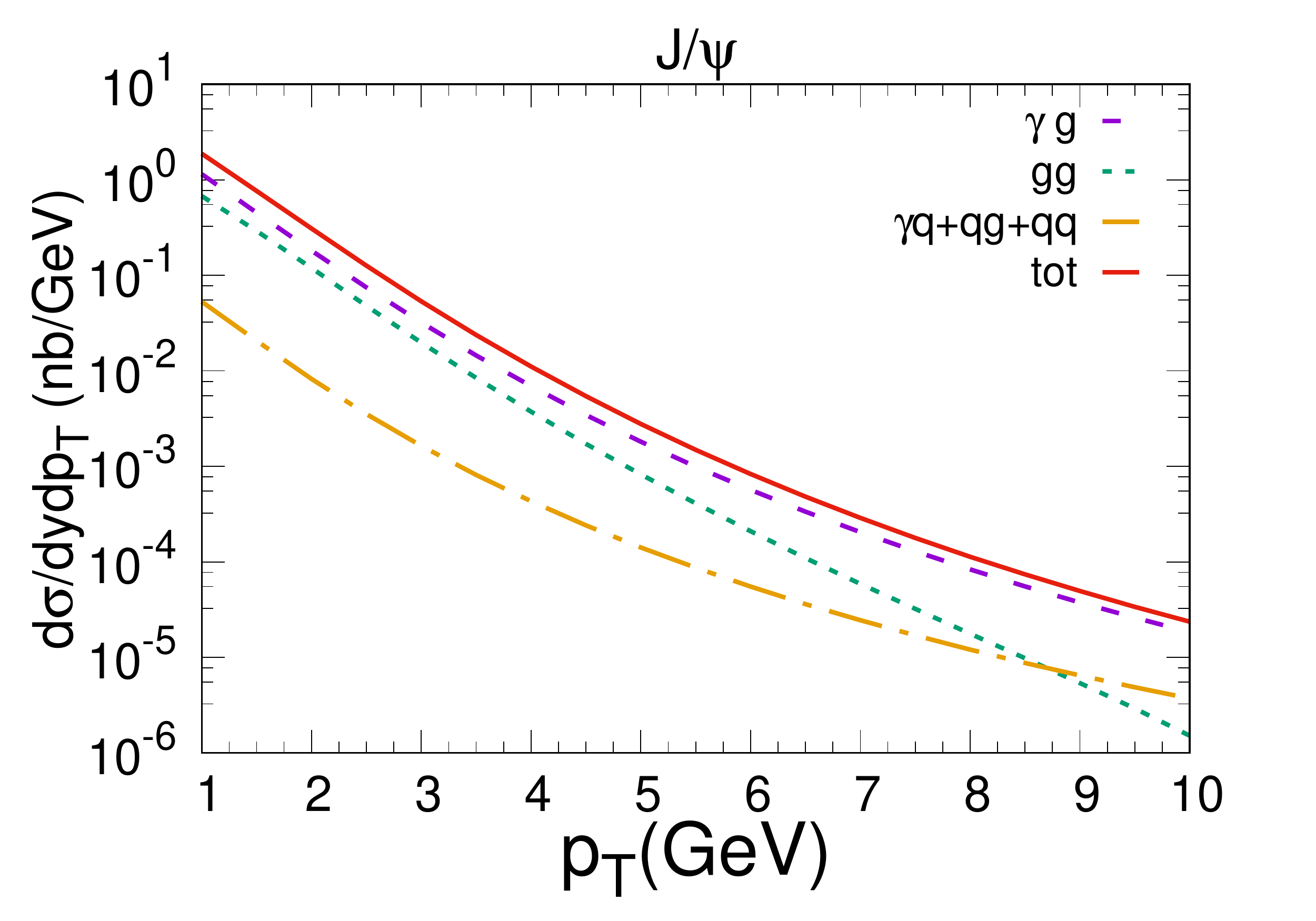, width=4cm,height=5cm,angle=270}
\epsfig{file=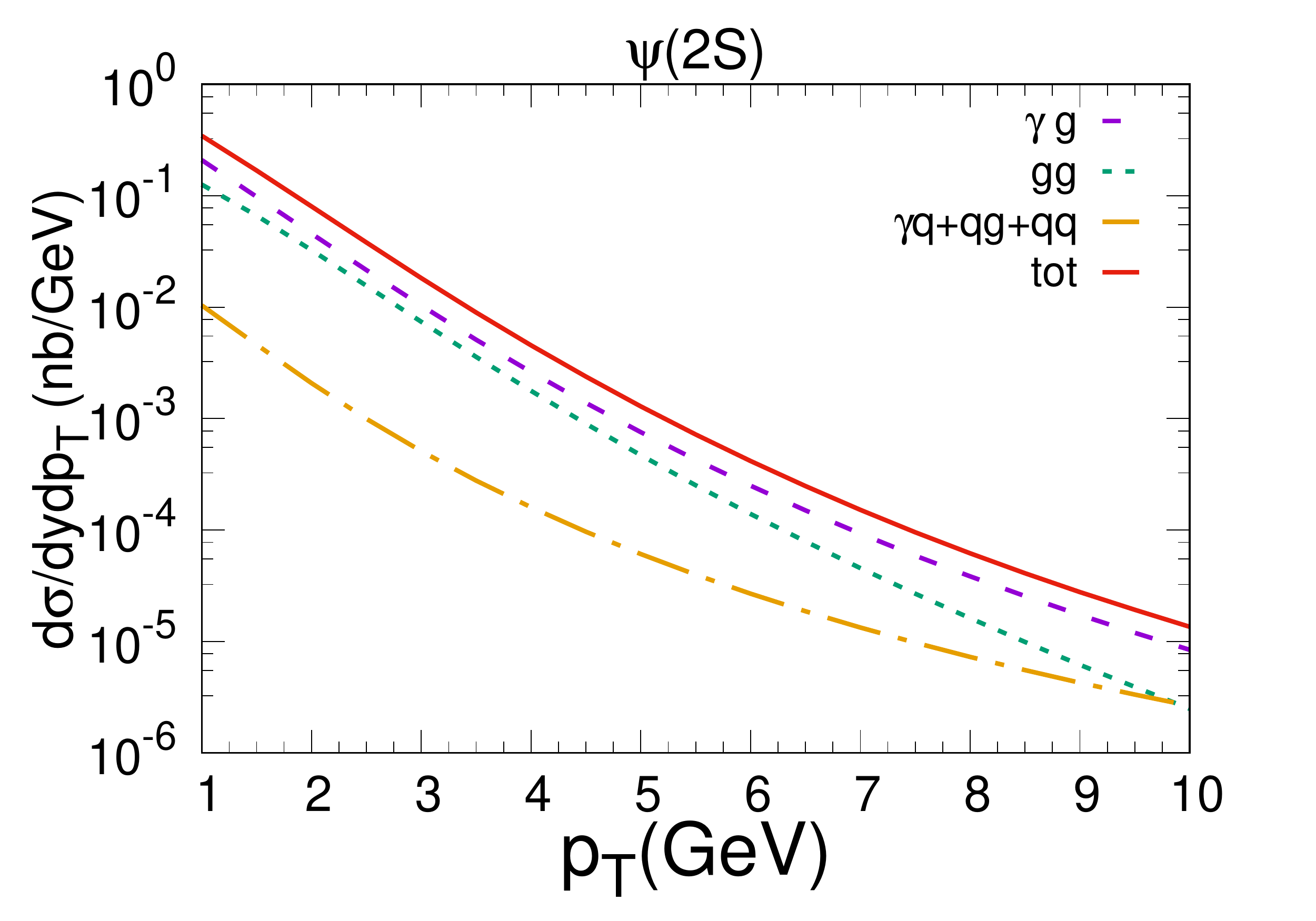, width=4cm,height=5cm,angle=270}
\epsfig{file=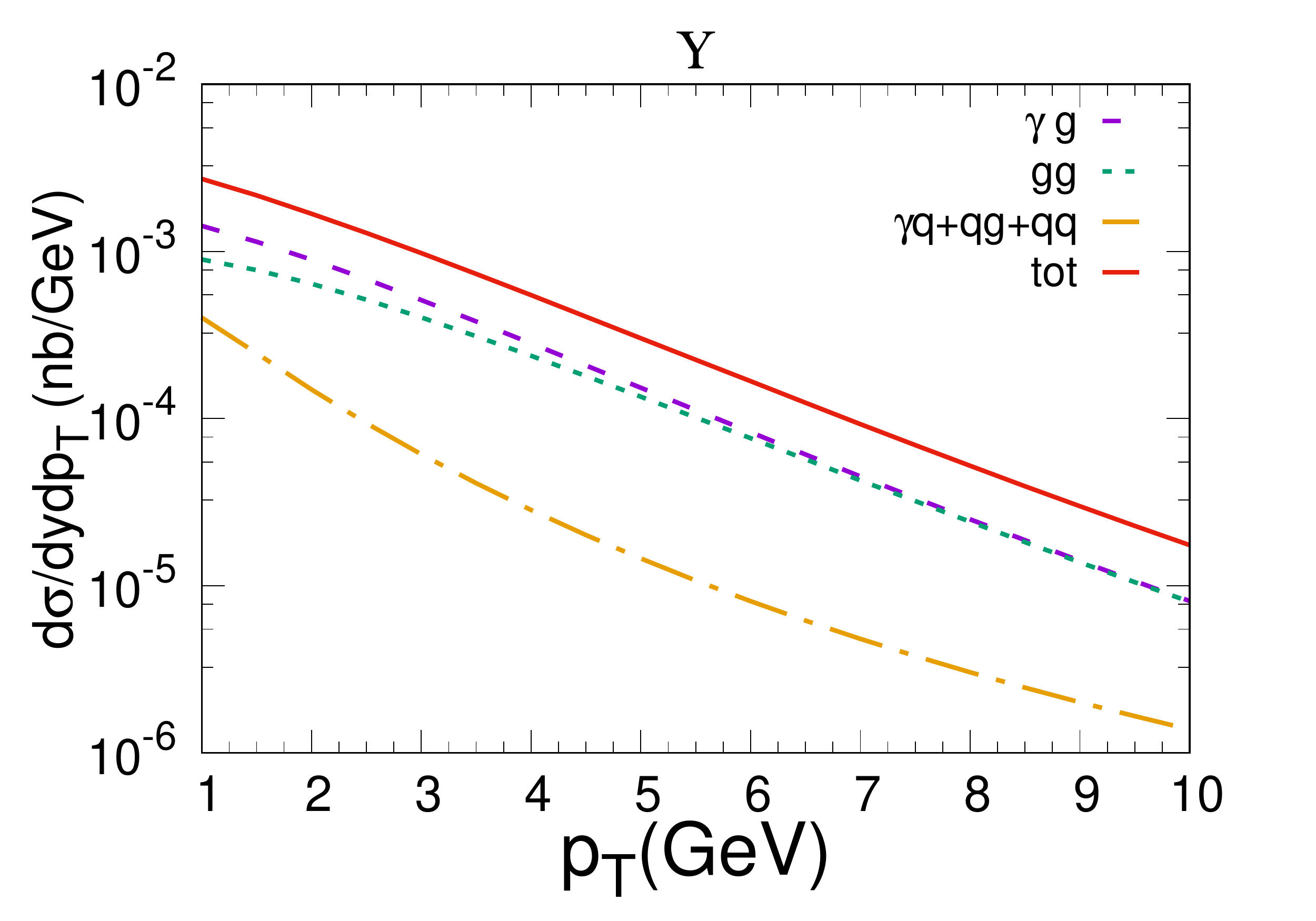, width=4cm,height=5cm,angle=270}
\end{center}
\caption{Transverse momentum distributions for the inclusive diffractive $J/\Psi$, $\Psi(2S)$ and $\Upsilon(1S)$ photoproduction at central rapidities($y = 0$) in $pp$ collisions at $\sqrt{s}$ = 5.02 TeV.}
\label{fig7}
\end{figure}

\begin{figure}[t]
\setlength{\unitlength}{1.5cm}
\begin{center}
\epsfig{file=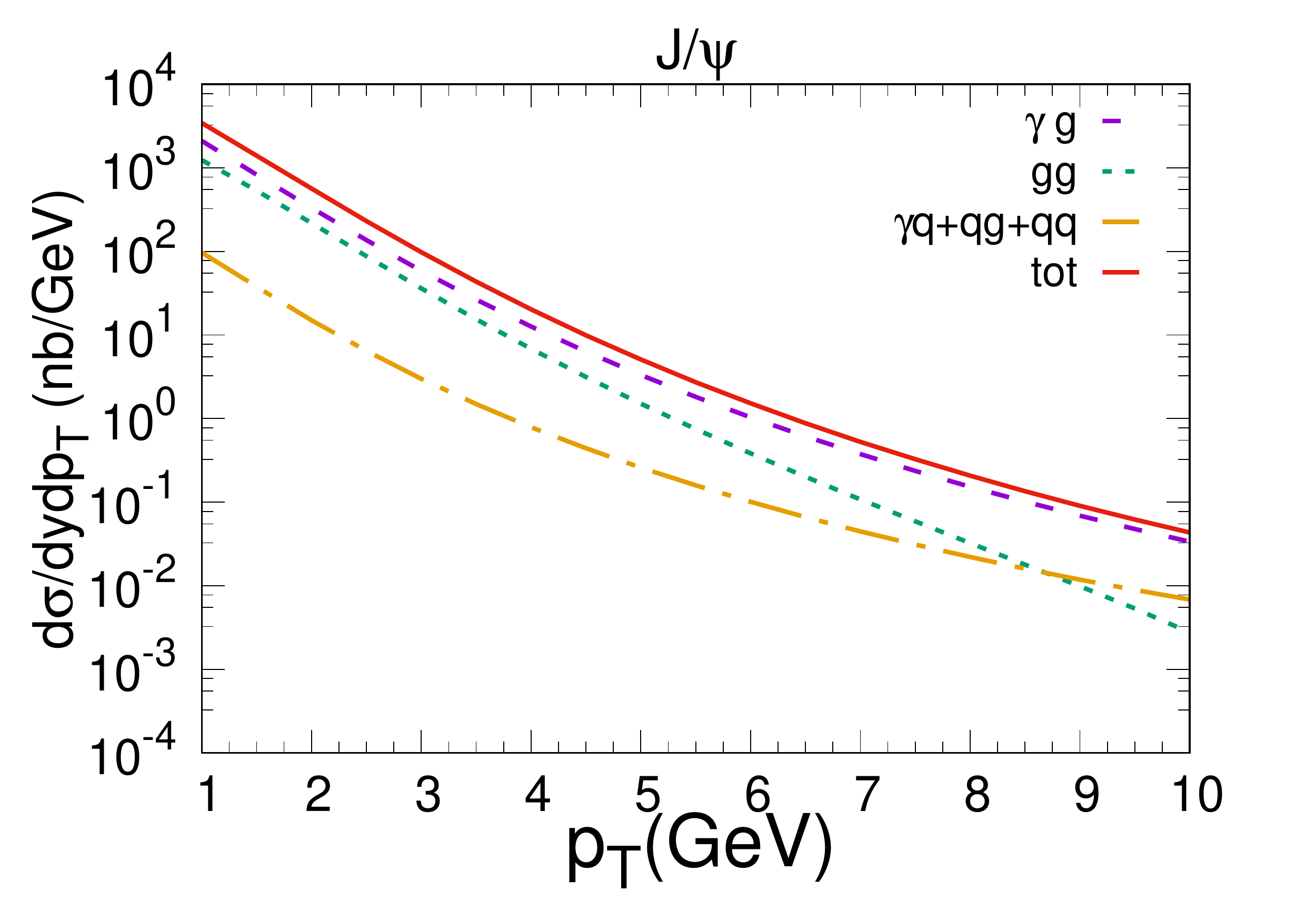, width=4cm,height=5cm,angle=270}
\epsfig{file=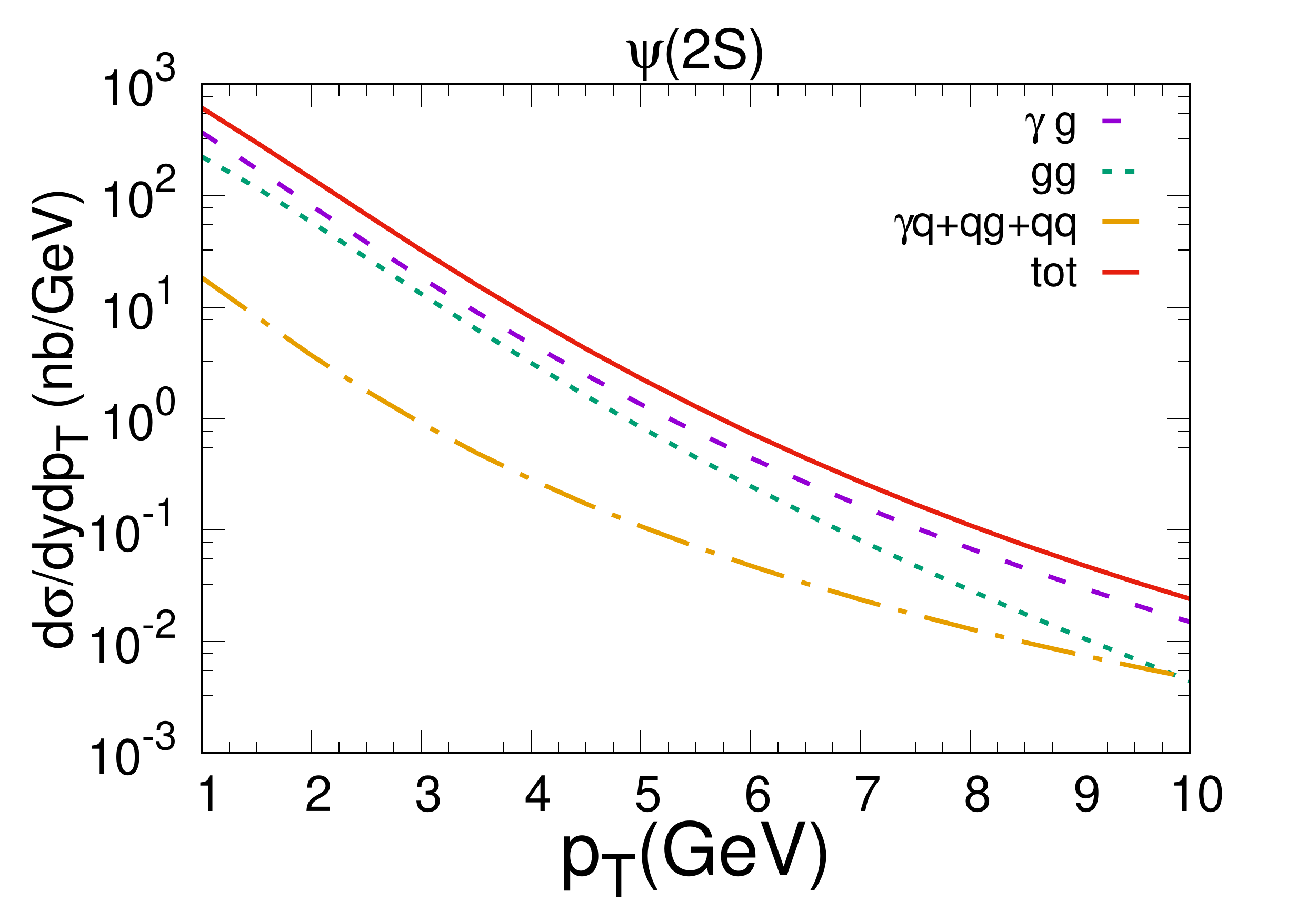, width=4cm,height=5cm,angle=270}
\epsfig{file=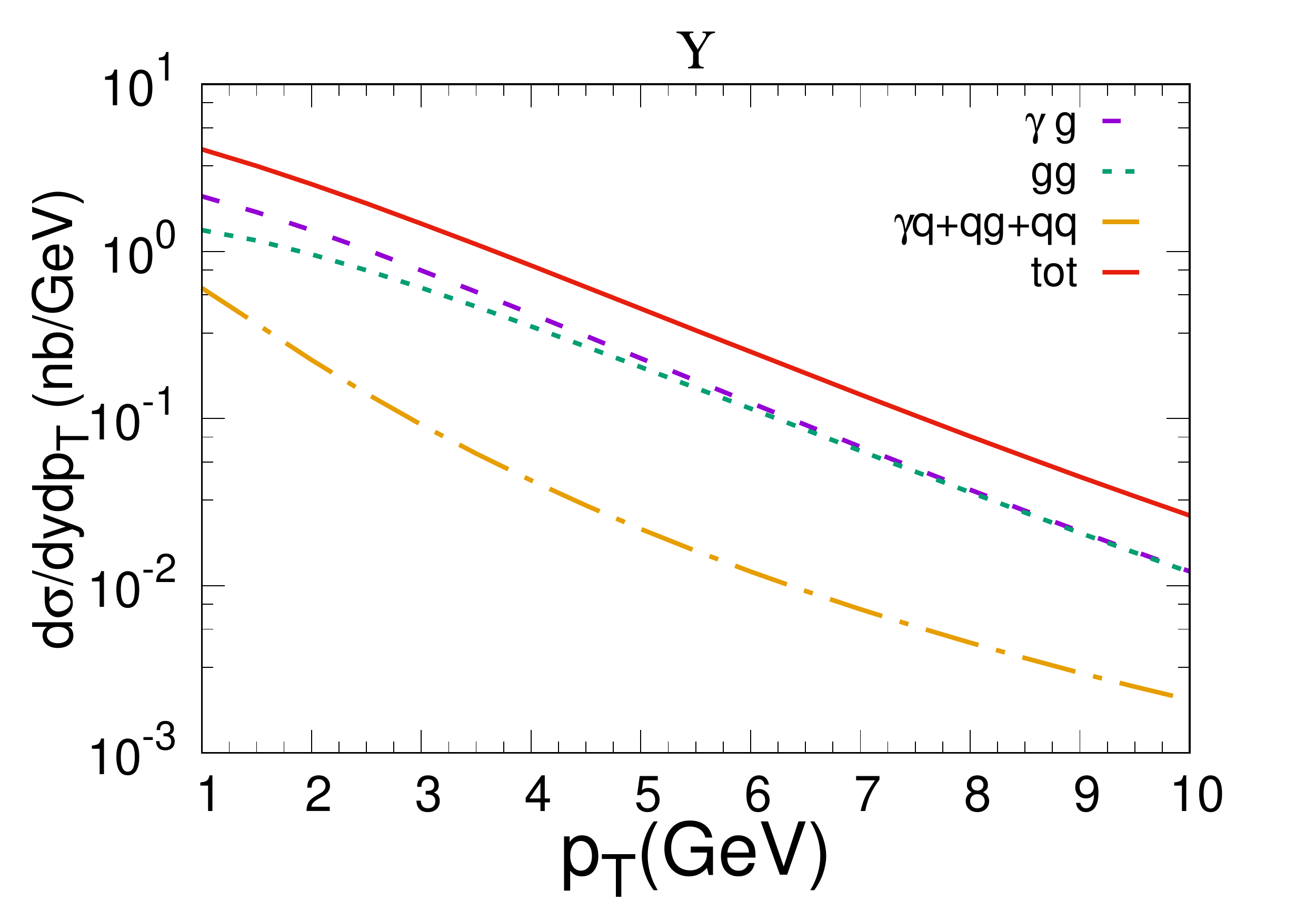, width=4cm,height=5cm,angle=270}
\end{center}
\caption{Transverse momentum distributions for the inclusive diffractive $J/\Psi$, $\Psi(2S)$ and $\Upsilon(1S)$ photoproduction at central rapidities($y = 0$) in $pPb$ collisions at $\sqrt{s}$ = 5.02 TeV.}
\label{fig8}
\end{figure}
\begin{figure}[t]
\setlength{\unitlength}{1.5cm}
\begin{center}
\epsfig{file=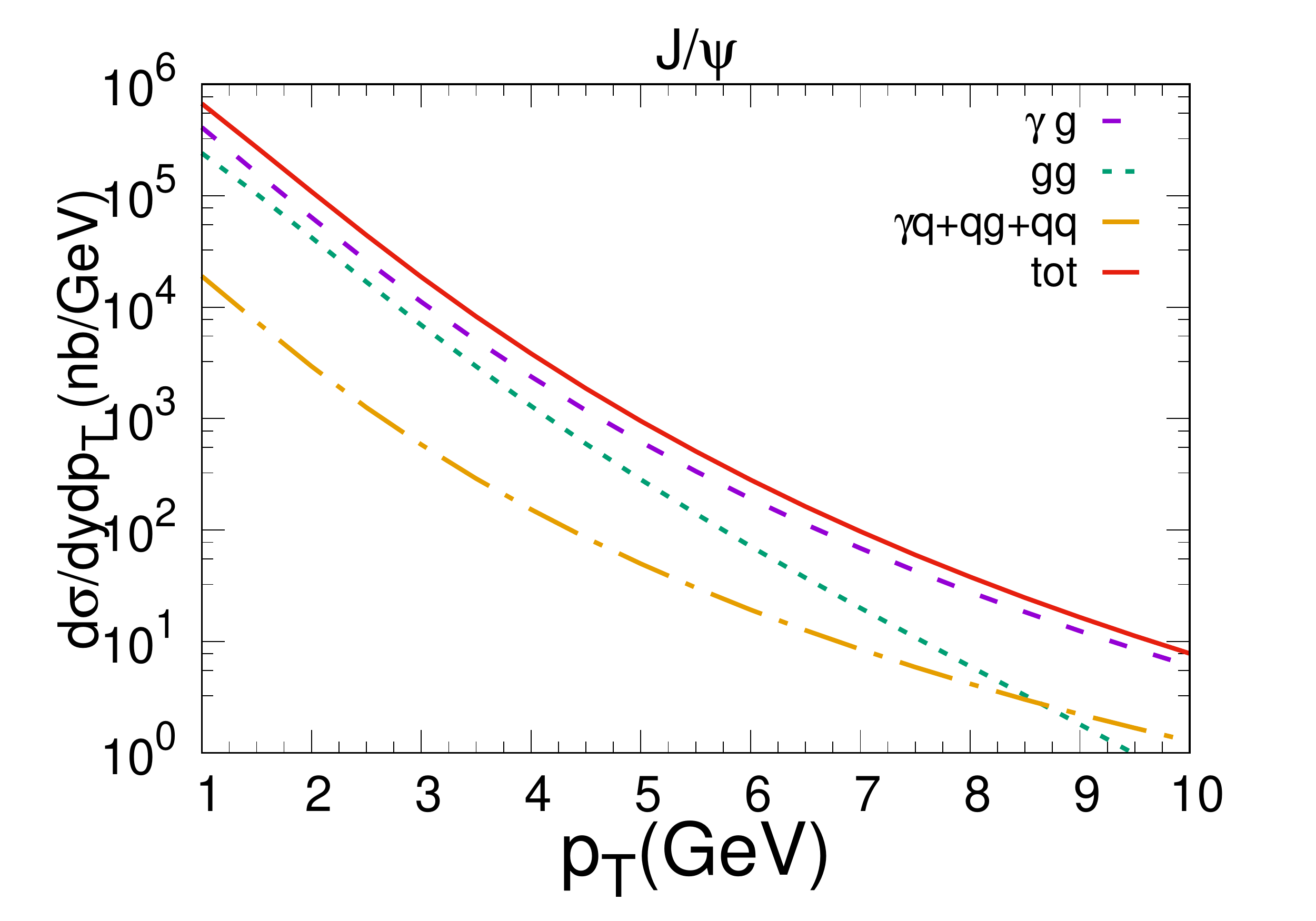, width=4cm,height=5cm,angle=270}
\epsfig{file=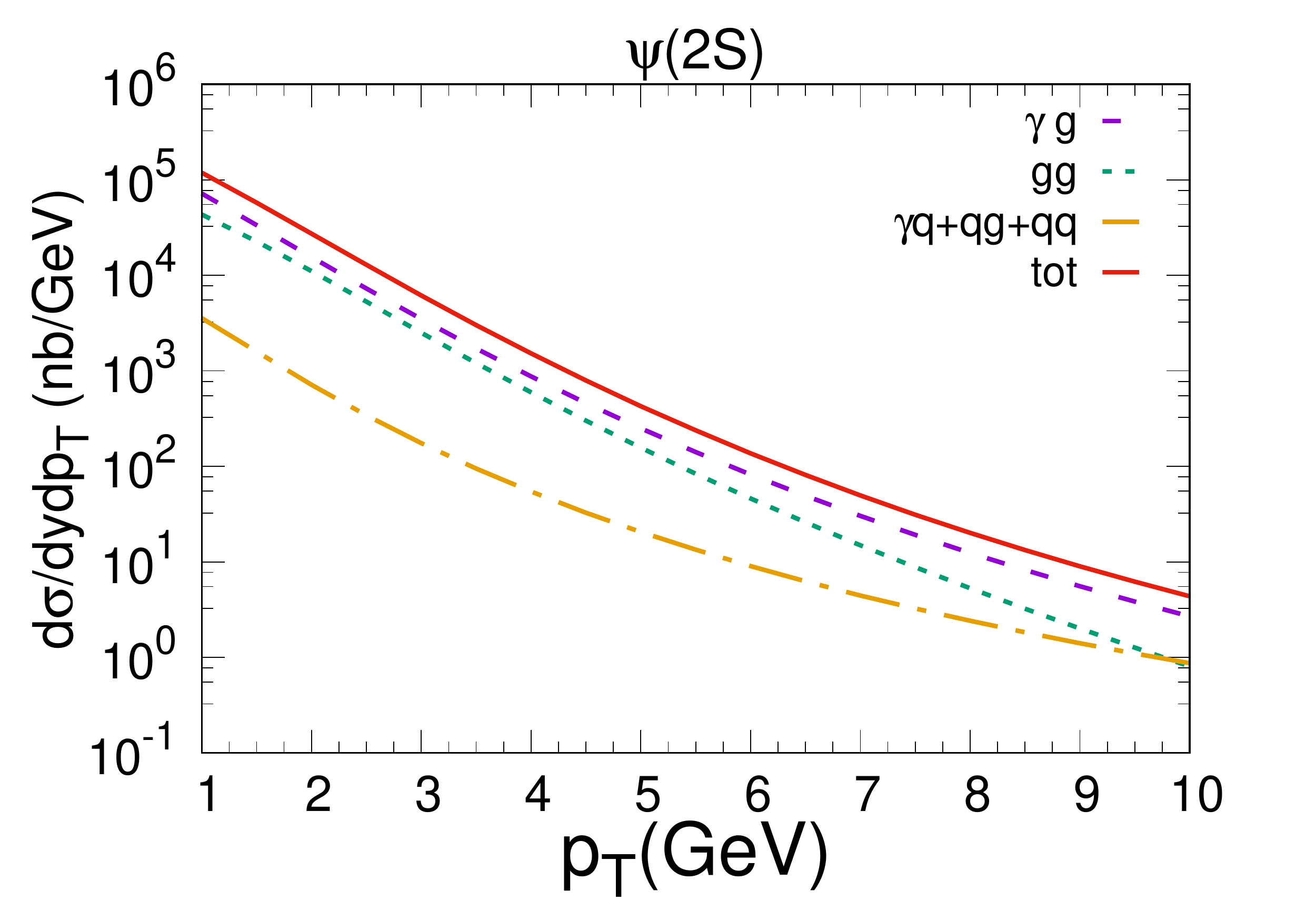, width=4cm,height=5cm,angle=270}
\epsfig{file=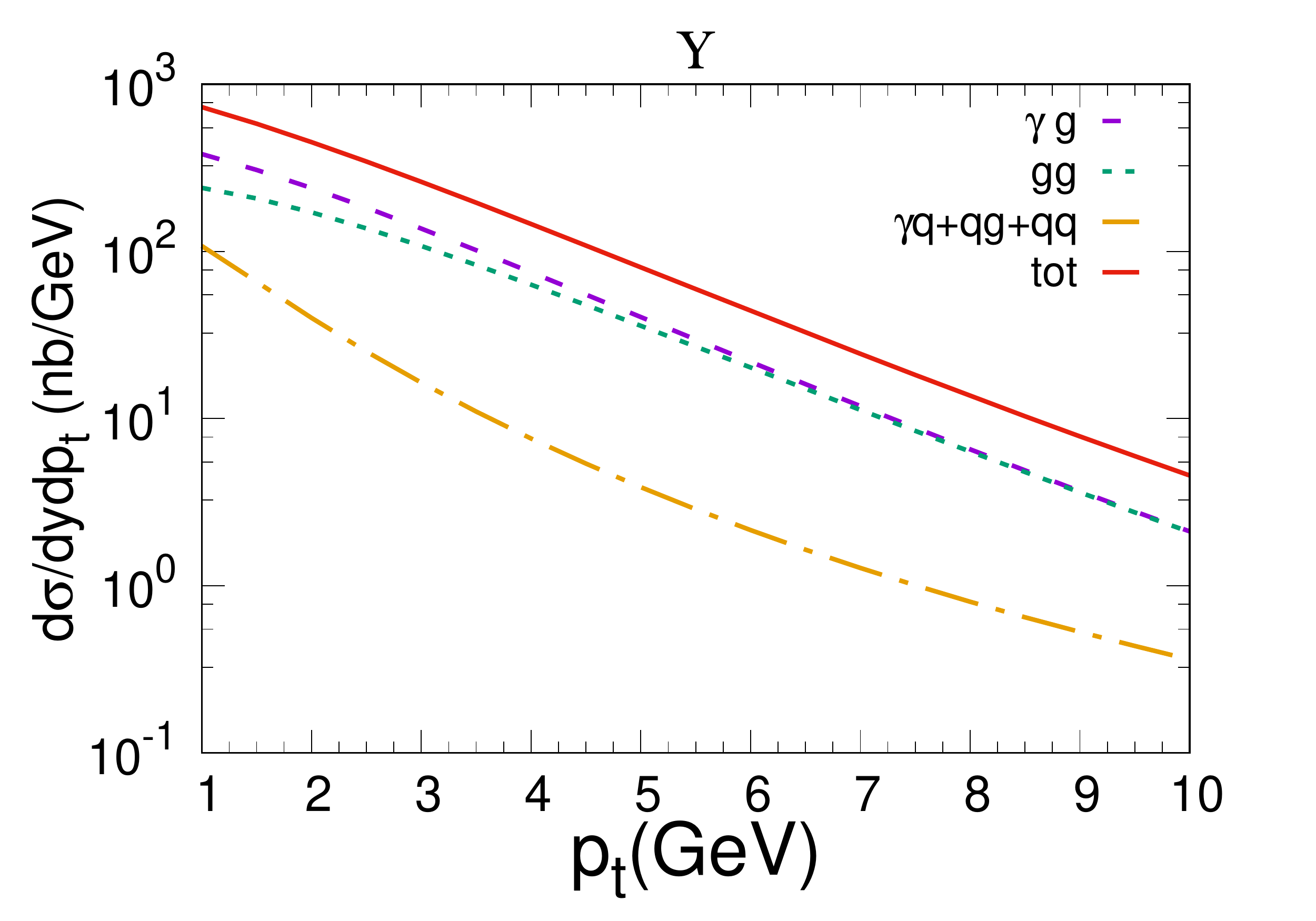, width=4cm,height=5cm,angle=270}
\end{center}
\caption{Transverse momentum distributions for the inclusive diffractive $J/\Psi$, $\Psi(2S)$ and $\Upsilon(1S)$ photoproduction at central rapidities($y = 0$) in $PbPb$ collisions at $\sqrt{s}$ = 5.02 TeV.}
\label{fig9}
\end{figure}

In Figs.\ref{fig7}, \ref{fig8} and \ref{fig9}, we present the predictions of the transverse momentum distribution for the inclusive diffractive $J/\Psi$, $\Psi(2S)$ and $\Upsilon(1S)$ photoproduction at central rapidities ($y$ = 0) in $pp$, $pPb$ and $PbPb$ collisions at $\sqrt{s} = 5.02$ TeV. As expected that the transverse momentum distributions of $\Upsilon(1S)$ are smaller than that of $J/\Psi$ and $\Psi(2S)$, due to the fact that the mass of $\Upsilon(1S)$ is larger than charmonium. Throughout the calculations, we find that the quark involved three subprocesses make a significant contributions to the heavy quarkounium inclusive diffractive transverse momentum distributions, since the percentage of the contributions from these three subprocesses can reach to $6\%$.


\section{Summary}
\label{sec:sum}
We extend our previous model, which is based on the NRQCD factorization formlism, to include quark involved three subprocesses to study the heavy quarkonium production. We call this approach as quark improved NRQCD model. The quark improved NRQCD model is use to calculate the total cross section of inclusive $J/\Psi$ production at HERA. By comparing our theoretical calculations of the $J/\Psi$ production with the experimental data, it finds that the contributions from the the three subprocesses mentioned above play a significant role in an accurate description of the heavy quarkonium production. Then we use the quark improved NRQCD model to study heavy quarkonium photoproduction at the LHC energies by combining with the resolved pomeron model. We make the predictions of the inclusive diffractive rapidity and transverse momentum distributions of $J/\Psi$, $\Psi(2S)$ and $\Upsilon(1S)$ in $pp$, $pPb$ and $PbPb$ collisions. We calculate the percentage of the quark involved three subprocesses in heavy quarkonium photoproduction, it finds that their contributions can reach to $8\%$ and $6\%$ at LHC energies, respectively. These results show that the quark involved three subprocesses may offer an insight to understand the underlying mechanism of heavy quarkonium production.


\begin{acknowledgments}
This work is supported by the National Natural Science Foundation of China under Grant Nos.11765005, 11305040, 11947119 and 11847152; the Fund of Science and Technology Department of Guizhou Province under Grant Nos.[2018]1023, and [2019]5653; the Education Department of Guizhou Province under Grant No.KY[2017]004; the National Key Research and Development Program of China under Grant No.2018YFE0104700, and Grant No.CCNU18ZDPY04.
\end{acknowledgments}


\end{document}